\documentclass[sigconf,natbib=true]{acmart}

\usepackage[T1]{fontenc}
\usepackage[latin1]{inputenc}
\usepackage{microtype}
\usepackage{xspace}
\usepackage{xcolor}
\usepackage{colortbl}
\usepackage{url}
\usepackage{float}
\usepackage{multirow}
\usepackage{arydshln}

\newcommand{\Ni}{(1)~}
\newcommand{\Nii}{(2)~}
\newcommand{\Niii}{(3)~}
\newcommand{\Niv}{(4)~}
\newcommand{\Nv}{(5)~}
\newcommand{\Na}{(a)~}
\newcommand{\Nb}{(b)~}

\newcommand{\textttsmall}[1]{\texttt{\small #1}}
\newcommand{\query}[1]{\textttsmall{#1}}
\newcommand{\domain}[1]{\textttsmall{\href{http://#1}{#1}}}

\newcommand{\cld}{\textttsmall{\href{https://github.com/google/cld3}{cld3}}\xspace}

\DeclareGraphicsExtensions{.pdf,.ai,.jpg,.png}
\setkeys{Gin}{pagebox=artbox}

\graphicspath{{sigir23-archive-query-log-figures/}}

\copyrightyear{2023}
\acmYear{2023}
\setcopyright{acmlicensed}
\acmConference[SIGIR '23]{Proceedings of the 46th International ACM SIGIR Conference on Research and Development in Information Retrieval}{July 23--27, 2023}{Taipei, Taiwan}
\acmBooktitle{Proceedings of the 46th International ACM SIGIR Conference on Research and Development in Information Retrieval (SIGIR '23), July 23--27, 2023, Taipei, Taiwan}
\acmPrice{15.00}
\acmDOI{10.1145/3539618.3591890}
\acmISBN{978-1-4503-9408-6/23/07}

\begin{document}

\title{The Archive Query Log: Mining Millions of Search Result Pages of Hundreds of Search Engines from 25 Years of Web Archives}

\author[J.\ H.\ Reimer]{Jan Heinrich Reimer}
\affiliation{
\institution{Friedrich-Schiller-Universit{\"a}t Jena}
\country{}
\city{}
}

\author[S.\ Schmidt]{Sebastian Schmidt}
\affiliation{
\institution{Leipzig University}
\country{}
\city{}
}

\author[M.\ Fr{\"o}be]{Maik Fr{\"o}be}
\affiliation{
\institution{Friedrich-Schiller-Universit{\"a}t Jena}
\country{}
\city{}
}

\author[L.\ Gienapp]{Lukas Gienapp}
\affiliation{
\institution{Leipzig University and ScaDS.AI}
\country{}
\city{}
}

\author[H.\ Scells]{Harrisen Scells}
\affiliation{
\institution{Leipzig University}
\country{}
\city{}
}

\author[B.\ Stein]{Benno Stein}
\affiliation{
\institution{Bauhaus-Universit{\"a}t Weimar}
\country{}
\city{}
}

\author[M.\ Hagen]{Matthias Hagen}
\affiliation{
\institution{Friedrich-Schiller-Universit{\"a}t Jena}
\country{}
\city{}
}

\author[M.\ Potthast]{Martin Potthast}
\affiliation{
\institution{Leipzig University and ScaDS.AI}
\country{}
\city{}
}

\renewcommand{\shortauthors}{Jan Heinrich Reimer et al.}

\begin{abstract}
The Archive Query Log~(AQL) is a previously unused, comprehensive query log collected at the Internet Archive over the last 25~years. Its first version includes 356~million queries, 137~million search result pages, and 1.4~billion search results across 550~search providers. Although many query logs have been studied in the literature, the search providers that own them generally do not publish their logs to protect user privacy and vital business data. Of the few query logs publicly available, none combines size, scope, and diversity. The~AQL is the first to do so, enabling research on new retrieval models and (diachronic) search engine analyses. Provided in a privacy-preserving manner, it promotes open research as well as more transparency and accountability in the search industry.
\end{abstract}

\keywords{query log, search engine result page, information retrieval history}

\ccsdesc[500]{Information systems~Query log analysis}

\maketitle

\section{Introduction}

Search engine query logs are a rich resource for many information retrieval applications~\cite{agosti:2012}, such as user behavior and user experience analysis, or query suggestions and query reformulations. When a query log also includes users' clicks and dwell time on search results, this is a valuable source of implicit relevance feedback about their information needs. Modern search engines use this feedback to train retrieval models for re-ranking~\cite{oosterhuis:2018,zhuang:2020}. However, query logs are also highly sensitive data that affect a number of stakeholders~\cite{grimmelmann:2007,koerber:2015}: First and foremost are user privacy concerns. Over time, if a user frequently uses a search engine, their query log can be enough to personally identify them and reveal a lot about their state of mind and health. To some extent, this also applies to persons or organizations mentioned or implied in queries or search results. Not least, relevance feedback from a query log is an important asset for search providers, commercial or otherwise.

The aforementioned arguments present valid concerns for refraining from publishing query logs. An additional, albeit debatable, rationale for major search providers stems from the expectations of increased transparency and accountability from governments, civil societies, affected users, and third parties, owing to their substantial market presence. Granting access to query logs would facilitate large-scale, independent investigations into the accuracy and fairness of their search results~\cite{grimmelmann:2007}, foster competition~\cite{argenton:2012}, support law enforcement efforts~\cite{cooper:2008}, and advance public information retrieval research. However, these objectives may not align with the best interests of search providers themselves.

\begin{table}[t]
\small
\centering
\renewcommand{\tabcolsep}{5.5pt}
\caption{The Archive Query Log~2022 (AQL-22) at a glance.}
\label{table-archive-query-log-top-10}
\vskip-2ex
\begin{tabular}{@{}llrrrrr@{}}
\toprule
  \multicolumn{2}{@{}l@{}}{\bfseries Search provider} &
  \multicolumn{1}{@{}c@{}}{\phantom{9,9}\bfseries URLs} &
  \multicolumn{1}{@{}c@{}}{\phantom{9}\bfseries Queries} &
  \multicolumn{1}{@{}c@{}}{\bfseries Queries} &
  \multicolumn{1}{@{}c@{}}{\phantom{9}\bfseries SERPs} &
  \multicolumn{1}{@{}c@{}}{\phantom{9,}\bfseries Results}
\\[-0.75ex]
  \multicolumn{2}{@{}l@{}}{\footnotesize (known domains)} &
  \multicolumn{1}{@{}c@{}}{\phantom{9,9}\footnotesize (total)} &
  \multicolumn{1}{@{}c@{}}{\phantom{9}\footnotesize (total)} &
  \multicolumn{1}{@{}c@{}}{\footnotesize (unique)} &
  \multicolumn{1}{@{}c@{}}{\phantom{9}\footnotesize (estimate)} &
  \multicolumn{1}{@{}c@{}}{\phantom{9,}\footnotesize (estimate)}
\\[-0.5ex]
\midrule
  \raisebox{-0.25ex}{\includegraphics[width=2ex]{logo-google}    } & Google     &    89.4\,M  &    72.7\,M  &   20.0\,M  &    28.0\,M  &    223.1\,M \\
  \raisebox{-0.25ex}{\includegraphics[width=2ex]{logo-youtube}   } & YouTube    &    41.8\,M  &    41.4\,M  &   11.3\,M  &    15.9\,M  &    339.2\,M \\
  \raisebox{-0.25ex}{\includegraphics[width=2ex]{logo-baidu}     } & Baidu      &    78.5\,M  &    69.6\,M  &    2.9\,M  &    26.8\,M  &    107.6\,M \\
  \raisebox{-0.25ex}{\includegraphics[width=2ex]{logo-qq}        } & QQ         &     0.5\,M  &     0.5\,M  &    0.1\,M  &     0.2\,M  &      2.1\,M \\
  \raisebox{-0.25ex}{\includegraphics[width=2ex]{logo-facebook}  } & Facebook   &     3.1\,M  &     0.2\,M  &    0.0\,M  &     0.1\,M  &      0.7\,M \\
  \raisebox{-0.25ex}{\includegraphics[width=2ex]{logo-yahoo}     } & Yahoo!     &     8.8\,M  &     2.8\,M  &    1.2\,M  &     1.1\,M  &      9.2\,M \\
  \raisebox{-0.25ex}{\includegraphics[width=2ex]{logo-amazon}    } & Amazon     &    66.8\,M  &     0.8\,M  &    0.3\,M  &     0.3\,M  &      7.8\,M \\
  \raisebox{-0.25ex}{\includegraphics[width=2ex]{logo-wikipedia} } & Wikipedia  &    68.5\,M  &     1.7\,M  &    0.6\,M  &     0.7\,M  &      7.0\,M \\
  \raisebox{-0.25ex}{\includegraphics[width=2ex]{logo-jd-dot-com}} & JD.com     &     4.4\,M  &     3.9\,M  &    0.4\,M  &     1.5\,M  &     16.0\,M \\
  \raisebox{-0.25ex}{\includegraphics[width=2ex]{logo-360}       } & 360        &     1.5\,M  &     1.1\,M  &    0.1\,M  &     0.4\,M  &      3.5\,M \\
\midrule
  \hspace{0.4em}\rotatebox{90}{...}                                & 540 others &   646.8\,M  &   161.8\,M  &   27.8\,M  &    62.4\,M  &    693.9\,M \\
\midrule
  $\sum$                                                           & 550        &  1010.2\,M  &   356.5\,M  &   64.5\,M  &   137.3\,M  &   1410.0\,M \\
\bottomrule
\end{tabular}
\vskip-2ex
\end{table}

\newcommand{\logname}[1]{{\scriptsize\color{gray}\xspace(#1)}} 
\newcommand{\bslink}[1]{{\href{#1}{\raisebox{-0.5ex}{\includegraphics[height=2ex]{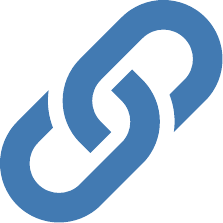}}}}}
\newcommand{\logurl}[1]{\hfill\scriptsize[{\color{blue!70}\href{#1}{url}}]}
\newcommand{\logmark}{}
\newcommand{\na}{{\color{gray}--}} 
\newcommand{\affi}{\raisebox{-0.25ex}{\includegraphics[width=.8em]{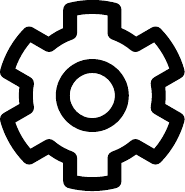}}\xspace} 
\newcommand{\affa}{\raisebox{0.1ex}{\includegraphics[width=.8em]{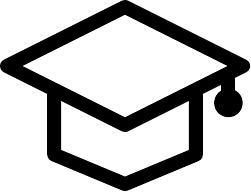}}\xspace} 
\newcommand{\affm}{\raisebox{0.1ex}{\includegraphics[width=.8em]{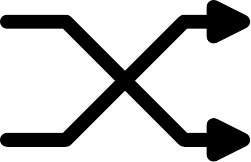}}\xspace} 

\begin{table*}[t]
\small
\renewcommand{\tabcolsep}{3.15pt}
\renewcommand{\arraystretch}{0.85}
\caption{
Overview of large-scale query logs used in previous work. 
Private logs are grouped by source. Each source is referenced by the paper using the largest sample. 
Number of usages is given under N\textsuperscript{\underline{o}}.
Timespan indicates crawled duration; year indicates date of last included query. 
Fields with `\na' are either not available or not specified. 
Languages are estimated.
The $\star$ marks logs still available for download; \protect\affi marks industry; \protect\affa academic; \protect\affm mixed affiliations.
}
\label{table-public-query-logs}%
\begin{tabular}{@{}c@{\hspace{.35em}}lrrrrrrrlllccr@{\ }r@{}}
\toprule
& \textbf{Source}\logname{name}
& \kern-6em{\scriptsize Clickable link:}\hspace{.25em}{\raisebox{-0.5ex}{\includegraphics[height=2ex]{icon-url.pdf}}} 
& \textbf{Queries}
& \kern-0.25em\textbf{Queries}\,(uniq.)\kern-1em
& \textbf{Sessions}
& \textbf{Users}
& \textbf{Clicks}
& \textbf{Results}
& \textbf{Task}
& \kern-0.5em\textbf{Lang.}\kern-0.6em
& \kern-0.6em\textbf{Span}\kern-0.8em
& \textbf{Year}
& \kern-0.6em\textbf{Aff.}\kern-0.6em
& \kern-0.6em\textbf{Ref.} 
& \kern.2em\textbf{N\textsuperscript{\underline{o}}} \\
\midrule
\multirow{26}{*}{\rotatebox{90}{\textbf{Private} (sample)}}
& AltaVista                                  &                                                                                                                                                           & 575,244,993 & 153,645,050 & \na        & 285,000,000 & \na        &           \na & Web   & en    & 1m  & 1998 & \affi & \cite{silverstein:1999} & 3  \\
& Dreamer                                    &                                                                                                                                                           & 2,184,256   & 228,566     & \na        & \na         & \na        &           \na & Web   & zh    & 3m  & 1998 & \affa & \cite{chuang:2003}      & 1  \\
& Excite                                     &                                                                                                                                                           & 51,473      & 18,098      & \na        & 18,113      & \na        &           \na & Web   & en    & \na & 1998 & \affa & \cite{jansen:1998}      & 3  \\
& Infoseek                                   &                                                                                                                                                           & 19,933,187  & \na         & \na        & \na         & \na        &           \na & Web   & zh    & 1m  & 1998 & \affm & \cite{lima:1999}        & 1  \\
& GAIS                                       &                                                                                                                                                           & 475,564     & 114,182     & \na        & \na         & \na        &           \na & Web   & zh    & 2w  & 1999 & \affa & \cite{chuang:2003}      & 1  \\
& Lycos                                      &                                                                                                                                                           & 500,000     & 243,595     & \na        & \na         & 500,000    &       361,906 & Web   & en    & 1d  & 2000 & \affm & \cite{beeferman:2000}   & 1  \\
& Yahoo!                                     &                                                                                                                                                           & 2,369,282   & \na         & \na        & \na         & 21,421     &           \na & Web   & en,zh & 4m  & 2000 & \affa & \cite{huang:2003}       & 9  \\
& OpenFind                                   &                                                                                                                                                           & 2,493,211   & \na         & \na        & \na         & \na        &           \na & Web   & en    & 1y  & 2000 & \affa & \cite{chuang:2003}      & 1  \\
& Encarta                                    &                                                                                                                                                           & 2,772,615   & \na         & 2,772,615  & \na         & \na        &           \na & Web   & en    & 1m  & 2002 & \affm & \cite{wen:2002}         & 1  \\
& Utah State Gov.                            &                                                                                                                                                           & 792,103     & 575,389     & 458,962    & 161,042     & 323,285    &           \na & Web   & en    & 5m  & 2003 & \affa & \cite{chau:2005}        & 1  \\
& Timway                                     &                                                                                                                                                           & 1,255,633   & \na         & \na        & \na         & \na        &           \na & Web   & zh    & 3m  & 2004 & \affa & \cite{lu:2006}          & 1  \\
& MetaSpy                                    &                                                                                                                                                           & 580,000     & \na         & \na        & \na         & \na        &           \na & Web   & en    & 5d  & 2005 & \affa & \cite{ahmad:2005}       & 1  \\
& arXiv                                      &                                                                                                                                                           & 44,399      & \na         & \na        & 13,304      & 48,976     &           \na & Edu   & en    & 3m  & 2006 & \affa & \cite{radlinski:2006}   & 1  \\
& TodoCL                                     &                                                                                                                                                           & 192,924     & \na         & 348,035    & \na         & 360,641    &       360,641 & Web   & en    & 9m  & 2006 & \affm & \cite{dupret:2006}      & 1  \\
& kunstmuseum.nl                             &                                                                                                                                                           & 7,531       & 1,183       & \na        & \na         & \na        &           \na & Lib.  & nl    & 2y  & 2007 & \affa & \cite{arampatzis:2007}  & 1  \\
& Microsoft AdCenter                         &                                                                                                                                                           & 27,922,224  & 27,922,224  & \na        & \na         & 7,820,000  &           \na & Ads   & en    & 2m  & 2007 & \affa & \cite{kuo:2016}         & 2  \\
& Europeana                                  &                                                                                                                                                           & 3,024,162   & 1,382,069   & \na        & \na         & \na        &           \na & Lib.  & en    & 6m  & 2011 & \affm & \cite{ceccarelli:2011}  & 1  \\
& GNU IFT                                    &                                                                                                                                                           & 2,099       & \na         & \na        & \na         & \na        &         4,754 & Img.  & en    & 1y  & 2011 & \affa & \cite{morrison:2011}    & 1  \\
& INDURE                                     &                                                                                                                                                           & 14,503      & 2,923       & 85,215     & 6,434       & \na        &           \na & Edu   & en    & 3m  & 2011 & \affa & \cite{fang:2011}        & 1  \\
& Bing Videos                                &                                                                                                                                                           & 1,218,936   & 445,859     & 174,955    & 174,955     & \na        &           \na & Vid.  & en    & 1w  & 2011 & \affm & \cite{kofler:2012}      & 1  \\
& Baidu                                      &                                                                                                                                                           & 362,994,092 & 10,413,491  & 87,744,130 & \na         & \na        &    13,126,252 & Web   & zh    & \na & 2012 & \affi & \cite{zhao:2012}        & 4  \\
& CADAL Library                              &                                                                                                                                                           & 45,892      & \na         & 81,759     & \na         & \na        &       164,822 & Lib.  & en    & \na & 2012 & \affa & \cite{yi:2012}          & 1  \\
& Taobao                                     &                                                                                                                                                           & 1,410,960   & \na         & \na        & 4,285       & \na        &           \na & Prod. & zh    & 1m  & 2013 & \affa & \cite{zhou:2017}        & 2  \\
& Startpagina                                &                                                                                                                                                           & 10,000,000  & \na         & \na        & \na         & \na        &           \na & Web   & nl    & 1m  & 2014 & \affm & \cite{guijt:2015}       & 1  \\
& parsijoo.ir                                &                                                                                                                                                           & 27,000,000  & \na         & \na        & \na         & \na        &           \na & Web   & fa    & 2y  & 2017 & \affa & \cite{mansouri:2018}    & 1  \\
& CiteSeerX                                  &                                                                                                                                                           & 78,124,884  & 14,759,852  & \na        & \na         & \na        &           \na & Edu   & en    & 4y  & 2021 & \affa & \cite{rohagti:2021}     & 1  \\
\addlinespace[0.5ex]\hdashline\addlinespace[0.5ex]
\hspace{0.2em}\rotatebox{90}{...} & \multicolumn{6}{@{}l@{}}{15 query logs from undisclosed sources, all private, 5 \affa, 2 \affi, 8 \affm } & \multicolumn{8}{@{}r@{\hspace{.225em}}}{\cite{liao:2011,huang:2003,gu:2011,jiang:2013,baeza-yates:1993,beitzel:2004,cao:2010,baeza-yates:2015,radlinski:2011,yao:2010,sun:2010,huang:2013,ling:2001,hu:2012,jiang:2016}}      & \na\\
\midrule                                                        
\multirow{14}{*}{\rotatebox{90}{\textbf{Public} (exhaustive)}}
& PubMed\logmark                             &                                                                                                                                                           & 2,996,301   & \na         & \na        & 627,455     & \na        &           \na & Med.  & en    & 1d  & 2005 & \affa & \cite{herskovic:2007}   & 3  \\
& AOL\logmark\logname{AOL Query Log '06}     & \bslink{https://web.archive.org/web/20130609163138/http://www.cim.mcgill.ca/~dudek/206/Logs/AOL-user-ct-collection/U500k_README.txt}                      & 36,389,567  & 10,154,742  & \na        & 657,426     & 19,442,629 &    19,442,629 & Web   & en    & 3m  & 2006 & \affi & \cite{pass:2006}        & 14 \\
& MSN\logmark\logname{MS RFP'06}             & \bslink{https://web.archive.org/web/20070203002037/http://research.microsoft.com/ur/us/fundingopps/RFPs/Search_2006_RFP.aspx}                             & 14,921,285  & \na         & 14,921,285 & \na         & \na        &           \na & Web   & en    & 1m  & 2006 & \affi &                         & 17 \\
& Belga News Agency\logmark                  & \bslink{https://web.archive.org/web/20090625005923/http://retrieve.shef.ac.uk/~imageclef/}                                                                & 1,402,990   & \na         & \na        & \na         & 5,697,287  &       498,039 & Img.  & en    & 1y  & 2008 & \affa & \cite{morrison:2013}    & 2  \\
& bildungsserver.de\logmark\logname{DBMS}    & \bslink{https://web.archive.org/web/20110904134728/http://www.uni-hildesheim.de/logclef/Daten/DBS_file_descrption.pdf}                                    & 98,512      & 31,347      & 65,513     & \na         & 68,604     &           \na & Edu.  & de    & \na & 2009 & \affa & \cite{mandl:2010}       & 4  \\
& Gov2 Crawl\logname{LETOR 4.0}\logmark      & \bslink{https://web.archive.org/web/20230220183106/https://www.microsoft.com/en-us/research/project/letor-learning-rank-information-retrieval/letor-4-0/} & 2,500       & \na         & \na        & \na         & \na        &           \na & Web   & en    & \na & 2009 & \affi & \cite{qin:2013}         & 2  \\
& Sogou\logmark                              & \bslink{https://web.archive.org/web/20190923175811/http://www.sogou.com/labs/resource/q.php}                                                              & 18,393,652  & 4,580,463   & \na        & 8,168,051   & \na        &    14,075,717 & Web   & zh    & 1m  & 2009 & \affa & \cite{zhang:2009}       & 4  \\
& Tumba!                                     &                                                                                                                                                           & 458,623     & \na         & \na        & \na         & \na        &           \na & Web   & en    & \na & 2009 & \affa & \cite{mandl:2009}       & 1  \\
& European Library\logmark\logname{TEL}      & \bslink{https://web.archive.org/web/20110627163614/http://www.uni-hildesheim.de/logclef/Daten/LogCLEF2009_file_description.pdf}                           & 162,642     & \na         & 75,100     & \na         & \na        &           \na & Lib.  & en    & 1y  & 2010 & \affa & \cite{mandl:2009}       & 4  \\
& Yandex\logmark                             & \bslink{https://web.archive.org/web/20121108060407/http://switchdetect.yandex.ru/en/datasets}                                                             & 10,139,547  & \na         & \na        & 956,536     & \na        &    49,029,185 & Web   & ru    & \na & 2011 & \affi &                         & 2  \\
& Bing\logmark\logname{MS Image Ret. Ch.}    & \bslink{https://web.archive.org/web/20131124064042/http://web-ngram.research.microsoft.com/GrandChallenge/Datasets.aspx}                                  & 11,701,890  & \na         & \na        & \na         & \na        &           \na & Img.  & en    & \na & 2013 & \affi &                         & 1  \\
& Bing\logmark\logname{ORCAS}                & \bslink{https://microsoft.github.io/msmarco/ORCAS.html}$^\star$\kern-0.6em                                                                                & 18,823,602  & \na         & \na        & \na         & 18,823,602 &    18,823,602 & Web   & en    & \na & 2020 & \affm & \cite{craswell:2020}    & 3  \\
& AOL\logmark\logname{AOLIA}                 & \bslink{https://github.com/terrierteam/aolia-tools}$^\star$\kern-0.6em                                                                                    & 11,337,160  & \na         & \na        & \na         & \na        &     1,525,524 & Web   & en    & \na & 2022 & \affa & \cite{macavaney:2022}   & 1  \\
& StackOverflow\logmark                      & \bslink{https://archive.org/details/stackexchange}$^\star$\kern-0.6em                                                                                     & 9,046,179   & \na         & 16,164,506 & \na         & \na        &           \na & Q\&A  & en    & \na & 2022 & \affi &                         & 2  \\
\addlinespace[0.5ex]\hdashline\addlinespace[0.5ex]
& \textbf{Archive Query Log}\logname{AQL}    & \bslink{https://www.tira.io/task/archive-query-log}$^\star$\kern-0.6em                                                                                    & 356,450,494 & 64,544,345  & \na        & \na         & \na        & 1,409,974,669 & Multi & Multi & 25y & 2022 & \affa &                         & \\
\bottomrule
\end{tabular}
\end{table*}

We have uncovered and acquired an extensive query log that has accumulated at the Internet Archive over the last 25~years. We call this new resource Archive Query Log~(AQL). Table~\ref{table-archive-query-log-top-10} gives an overview of the first version of~2022 and the top ten search providers as per fused snapshots of Alexa website traffic rankings. Shown are the respective numbers of archived~URLs, the queries extracted from them, archived search result pages~(SERPs), and results linked to them. The SERPs of many queries have been archived multiple times, enabling diachronic analyses. At the time of writing, we collect this data for a total of 550~search providers. The Archive Query Log~2022 includes 356~million queries (65~million unique), 137~million search result pages, and 1.4~billion search results---an unprecedented scale for a public query log. Based on a comprehensive review of public and private query logs used in the literature (Section~\ref{part2}), we detail our acquisition method (Section~\ref{part3}), initial analyses (Section~\ref{part4}), and discuss our plan to share the data with the information retrieval community in a privacy-preserving manner, as well as limitations and ethical considerations (Section~\ref{part5}).%
\footnote{AQL code: \url{https://github.com/webis-de/SIGIR-23}}\textsuperscript{\hspace*{-0.1em},}%
\footnote{AQL access: \url{https://tira.io/task/archive-query-log}}

\section{Background and Related Work}
\label{part2}

We systematically review the use of query logs and search result pages in information retrieval research, and briefly discuss search transparency and accountability as well as the Internet Archive.

\subsection{Query Logs}

Table~\ref{table-public-query-logs} compiles an overview of 14~public and 31~private query logs from a focused literature review. Using the DBLP%
\footnote{\url{https://dblp.org/}}~title search, we screened all publications that mention ``query log'', ``click log'', or ``clickthrough'' in their title---a high-precision heuristic to ensure logs play a role, at the expense of recall. From the 642~publications found, the 492~related to information retrieval (e.g.,~not databases) were downloaded. We then searched for occurrences of the pattern ``\,\textttsmall{<number> <qualifier>} queries\,'' in them with regular expressions, assuming that virtually all researchers using query logs also specify how large they are.%
\footnote{Examples: ``1 million queries'', ``386,879 queries'', ``386\,879 queries'', or ``386k queries''. A qualifier is a sequence of up to 20~characters excluding end of sentence punctuation.}
This facilitated the manual extraction of passages and tables from 120~randomly selected publications for Table~\ref{table-public-query-logs}. Some entries were added manually to cover public logs.

Despite the fact that query logs are rarely published, researchers in academia have sought alternative means of access, usually by collaborating with many search providers large and small. Weighted by the number of publications per log, research with very large query logs was conducted in industry and at major search providers. The AOL~log~\cite{pass:2006} is the only exception. Together with its recent extension AOLIA~\cite{macavaney:2022} they are the largest publicly available query logs. The AltaVista~log and the Baidu~log are the two largest private logs reported. Our Archive Query Log is on par with the latter two. The ratio of unique queries to all queries averages~0.24. Our log's ratio of~0.18 is slightly lower due to its multilingual nature. In addition to queries, organic search engine query logs may include information about users, clicks, sessions, and results, while our log only includes queries, SERPs, and the result documents themselves.

Given the main tasks for which query logs are used, the~AQL can be used to study many---though not all---of them at a scale not easily attainable for academic researchers:
Query understanding involves analyzing user information behavior. Subtasks include determining the user's search intent~\cite{jansen:2005a} and examining user populations~\cite{jansen:1998}, particularly with respect to geographic~\cite{spink:2002} and temporal~\cite{jansen:2005} dimensions. In addition, much research has focused exclusively on how people search for health information, from both consumer~\cite{palotti:2016} and expert perspectives~\cite{scells:2022}.
Query suggestion involves exploiting query logs to recommend alternative queries to the user. Subtasks include clustering~\cite{beeferman:2000}, query similarity~\cite{chien:2005}, and query expansion using relevance feedback~\cite{cui:2002,cui:2003,huang:2003}. The~AQL may support both tasks in general, in particular when used for pre-training. However, model transfer to a specific application domain will be required, as well as organic log data for practical use cases.

Session analysis examines how users reformulate their queries across one or more sessions~\cite{agosti:2007}, a key subtask being session detection~\cite{hagen:2013,gayo-avello:2009}.
User modeling involves analyzing logs to build models of user interaction. Subtasks include examining the distributions of query lengths and query terms~\cite{jansen:2000b,silverstein:1999,spink:2001,wolfram:2001}, relevance feedback mechanisms~\cite{spink:2000}, and what users consider relevant~\cite{jansen:2003}. The~AQL does not support these tasks; it lacks session or user data.

Learning to rank exploits query logs to derive effective ranking models. Subtasks include developing click models~\cite{joachims:2002,joachims:2005} and models that incorporate implicit feedback such as dwell time on pages~\cite{agichtein:2006,zhang:2002}. The~AQL supports this task despite the lack of click data. \citeauthor{craswell:2020}'s~\cite{craswell:2020} rationale for the design of MS~MARCO corroborates this claim. Here, only passages (judged for relevance) from the top-ranked documents returned by Bing for a query are included as ground truth for training, which has proven to be sufficient to yield effective retrieval models. The same is true for the~AQL, where the ranked results of third-party retrieval models encode the domain expertise and the implicit relevance feedback from query logs that the respective search providers incorporated into them.

\subsection{Search Engine Result Pages}

Search engine result pages~(SERPs) are how search engines present results to users in response to a query. SERPs for web search typically consist of a list of links to web pages ranked by their relevance to the user's query, along with additional information such as snippets, images, and other features designed to help users meet their information needs. SERPs have been studied in information retrieval research for many years to understand how users interact with them, how they can be improved, and how they can present information more effectively to better meet user needs. The~AQL contains the SERP of 39\,\% of its queries.

One area of SERP research has focused on understanding how users interact with search results. Researchers have used techniques such as eye-tracking~\cite{arapakis:2008,jimmy:2019a,jimmy:2019b} and brain monitoring~\cite{moshfeghi:2021} to study how users perceive SERPs. These studies have led to a deeper understanding of how to improve the presentation of results and the ranking algorithms used by search engines. Another area of SERP research is the study of their design and layout~\cite{oliveira:2023a,oliveira:2023b}. These longitudinal studies show how SERPs evolve in response to new technologies. The~AQL provides millions of archived SERPs which include all necessary assets for showing them in a browser, so that they can be used for user studies and offline experiments.

\subsection{Transparency and Accountability}

In November~2022, the Digital Market Act~\cite{eu:2022a} and the Digital Services Act~\cite{eu:2022b} came into force in the European Union. The former applies primarily to ``gatekeepers'' in digital markets, such as Google for the search market, the latter to all digital services that act as (information) ``intermediaries''. Both laws contain provisions that, among other things, require search providers to increase data privacy, transparency, and accountability, with the goal of ensuring fair and open digital markets. In particular, legislators are allowed to exercise regulatory and market investigation powers, which may include looking into the algorithms used. The~AQL complements these measures and also gives civilian initiatives the means to conduct independent investigations of search providers. Previous studies on search accountability raise the question of how to inform users about a search engine's retrieval algorithms to raise awareness of how they work~\cite{cozza:2019,laidlaw:2009,langer:2008} and to ensure unbiased results~\cite{galindo:2017,li:2022}. While we cannot consider all previous work in this context, a recent overview was provided at the FACTS-IR workshop~\cite{olteanu:2019} on fairness, accountability, confidentiality, transparency, and safety in information retrieval. In terms of both algorithm transparency and search engine accountability, archived search result pages are perhaps one of the best representations of a search engine's behavior at a given time, and archiving them on a large scale allows for corresponding post-hoc analyses.

\subsection{Internet Archive}

The Internet Archive is a nonprofit digital library that has grown to become the largest and most comprehensive digital library in the world since its inception in~1996. In addition to providing access to extensive archives of books, audio recordings, videos, images, and software, the Internet Archive's best-known service is probably the Wayback Machine, which provides a digital archive of the web.%
\footnote{\url{https://web.archive.org/}} 
At the time of writing, it contains 806~billion web pages. We believe that the~AQL accumulated both due to accidental crawling by their crawlers and intentional archiving by their users, since any user can request archiving of any publicly accessible~URL. AOLIA~\cite{macavaney:2022} extends the original AOL~log by providing links to archived versions of its search results, originally specified as~URLs only. However, AOLIA's use is limited since only the landing pages of the domains from which search results were originally found could be recovered.

\section{Mining the Archive Query Log}
\label{part3}

\begin{figure}
\includegraphics[width=\linewidth]{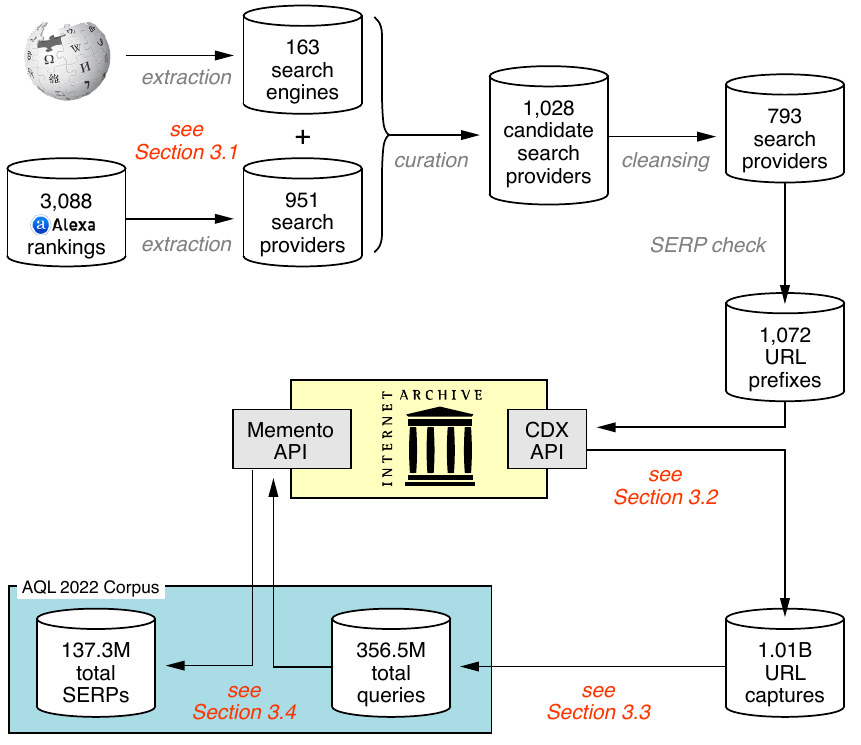}
\caption{Overview of the AQL creation process.}
\label{figure-aql-creation-overview}
\end{figure}

Besides general-purpose search engines, many other websites such as social media platforms offer a search function (a query field). The answer to a query is often encoded as URL linking to a SERP, which is displayed to the user. Like other URLs, these ``SERP URLs'' can be linked to by web pages and are thus included in automated web crawls. The Internet Archive, as the world's largest digital library of archived web pages, is likely to include many SERPs, a fact which can be exploited for large-scale query log mining. This section describes a multi-step process to mine a query log from the Internet Archive's Wayback Machine, which eventually becomes the Archive Query Log (see Figure~\ref{figure-aql-creation-overview}).

First, a list of popular search providers including general-purpose search engines and all kinds of media platforms is compiled (see upper part of Figure~\ref{figure-aql-creation-overview}; Section~\ref{provider-collection}).
Second, for each provider the second-level domains, known subdomains, and URL patterns under which SERPs are likely to be found are semi-automatically generated and the URL captures found in the Internet Archive (using the CDX API%
\footnote{\url{https://github.com/internetarchive/wayback/tree/master/wayback-cdx-server}}
of the Wayback Machine) are aggregated (Section~\ref{provider-domains-urls}).
Third, the queries are extracted from the URLs using provider-specific parsers  (Section~\ref{query-extraction}).
Fourth, the HTML content of archived SERPs is downloaded and parsed for metadata (URLs, titls, snippets, language, etc.; Section~\ref{serp-acquisition-and-parsing}). 
Both queries and metadata form the AQL-22 corpus (Section~\ref{aql-22}) which will be made accessible via the TIRA platform~\cite{potthast:2019p} as discussed in Section~\ref{part4}.

\subsection{Search Provider Collection}
\label{provider-collection}

Our search provider collection contains both
\Ni
websites that primarily act as search engines, and
\Nii
highly relevant websites that have been identified by their Alexa Rank.%
\footnote{The Alexa Rank was a ranking system that reflected the global popularity of website domains based on estimated visits; it was shut down end of~2022.}

Regarding~(1), we exploit a dedicated list of search engines on Wikipedia which we extend manually.%
\footnote{See \url{https://en.wikipedia.org/wiki/List_of_search_engines}}
Regarding~(2), we take the 3,088~archived snapshots of the Alexa top-1M ranking between June~2010 and November~2022%
\footnote{See  \url{https://web.archive.org/web/*/s3.amazonaws.com/alexa-static/top-1m.csv.zip}} 
and apply reciprocal rank fusion~\cite{cormack:2009} considering the 1,000~highest ranked website domains of each snapshot. The resulting list of 13,647~domains is narrowed down to 951~search providers by checking whether a search bar is present on the website's landing page of the respective provider. For this purpose we load the landing page directly or from an Internet Archive snapshot from~2022, render the page if JavaScript content is found, and check for HTML forms or \textttsmall{<div>}~elements containing the pattern ``search'' in its attributes.

The merged list of 1,028~unique candidate search providers is used to identify relevant URL patterns as well as suitable approaches for query extraction (see Section~\ref{provider-domains-urls}). Further manual curation steps weed out providers because they have been identified as spam, do not encode the search query in their URL, or offer only an autocomplete search that links directly to a page. Also, search providers are merged when more than one of their second-level domains appears in the merged Alexa list (e.g., \domain{so.com} is merged into \domain{360.com}). After these analyses and curation steps, the list includes 793~search providers.

\subsection{Provider Domains and URLs}
\label{provider-domains-urls}

Since many search providers are available under multiple second-level domains and/or subdomains,%
\footnote{Google lists 190~supported domains \url{https://www.google.com/supported_domains}}
we expand the list of the 793~provider domains manually and with publicly available lists.%
\footnote{E.g., \url{https://github.com/JamieFarrelly/Popular-Site-Subdomains}}  

However, on high-traffic domains, only a small fraction of all archived URLs is likely to link SERPs and thus is relevant for our purposes. To identify common prefixes of URLs that contain queries, we submit multiple test queries using the search provider's query field and examine the URL the request is redirected to. For discontinued or otherwise inaccessible websites, we resort to the most recent functional snapshot of the search provider's homepage in the Internet Archive. The final list of 1,072~URL prefixes is used to filter the list of available captures with the help of the Internet Archive's CDX~API. This API allows to request a list of URLs by the crawling date they were archived in the Wayback Machine for a certain domain or URL prefix. Via the CDX~API, a list of all available captures for each of the 1,072~URL prefixes is retrieved and filtered for successful captures with HTML content (i.e., HTTP status code~200). This process further narrows the search provider list with archived SERPs to~550.

Altogether, 1.1B~URL captures along with their crawling date are collected, an average of 1.8M~URLs per provider. Most of the captures originate from search engines (226M~URLs,~22\,\%) with Google being the largest contributor of archived URLs (89M~URLs,~9\,\%).

\subsection{Query Extraction}
\label{query-extraction}

To extract the query from a SERP URL, the URL is parsed into its components%
\footnote{RFC~2396; https://datatracker.ietf.org/doc/html/rfc2396.html}
and the query encoding is identified as one of three possible categories:
\Ni
URL parameter,
\Nii
path segment, or
\Niii
fragment identifier.
Examples of the first two patterns are illustrated in Figure~\ref{figure-url-decomposition}. 

For each of the three patterns, configurable query parsers are built using the \textttsmall{urllib} Python package:%
\footnote{\url{https://docs.python.org/3/library/urllib.html}}
\Ni
parsing a query parameter by its name (e.g., name \textttsmall{q} for the first URL in Figure~\ref{figure-url-decomposition}),
\Nii
parsing a segment from the URL's path component by its index (e.g., index~2 for the second URL in Figure~\ref{figure-url-decomposition}), or
\Niii
parsing a parameter from the fragment identifier by treating it like a query parameter.
In addition to the query, URLs can include a page number or offset. These help in reconstructing longer rankings from separate SERPs for the same query that are captured at nearly the same time. Google's SERPs, for instance, are paginated with 10~results per page. Thus, the page number can be used to infer the continued ranks of documents on the next page.

\begin{figure}[tb]
\begin{minipage}{\linewidth}
    \small
    \definecolor{urlignored}{RGB}{60,60,60}
    \definecolor{urlprefix}{RGB}{8,54,128}
    \definecolor{urlquery}{RGB}{204,0,0}
    \definecolor{urlpage}{RGB}{44,118,38}
    \newcommand{\urlignored}[1]{\textcolor{urlignored}{#1}}
    \newcommand{\urlprefix}[1]{\textcolor{urlprefix}{#1}}
    \newcommand{\urlquery}[1]{\textcolor{urlquery}{#1}}
    \newcommand{\urlpage}[1]{\textcolor{urlpage}{#1}}
    \newcommand{\urlrule}[0]{\rule[-1ex]{0.5pt}{4ex}}
    \renewcommand{\arraystretch}{1}
    \Na\textit{Query parameter:}\\[1.5ex]
    \begin{tabular}{@{}l@{}c@{}c@{}c@{}c@{}c@{}c@{}c@{}}
        \urlignored{scheme} & \urlignored{network loc.} & \urlignored{path} & \multicolumn{5}{c}{\urlignored{query string}} \\
        \arrayrulecolor{urlignored}%
        \cmidrule(r{0.25em}){1-1}%
        \cmidrule(l{0.25em}r{0.25em}){2-2}
        \cmidrule(l{0.25em}){3-3}
        \cmidrule{5-8} \\[-2.25ex]
        \texttt{\urlignored{https://}} & \texttt{\urlprefix{google.com}} & \texttt{\urlprefix{/search}} & \texttt{\urlignored{?}} & \texttt{\urlquery{q=\textbf{covid}+\textbf{19}+\textbf{usa}+\textbf{map}}} & \texttt{\urlignored{\&}} & \texttt{\urlpage{start=\textbf{10}}} & \texttt{\urlignored{\&ei={\footnotesize ...}}}\\[0.75ex]
        \arrayrulecolor{urlprefix}
        \cmidrule{2-3} \\[-4.2ex]
        \arrayrulecolor{urlquery}
        \cmidrule{5-5} \\[-4.25ex]
        \arrayrulecolor{urlpage}
        \cmidrule{7-7}
        & \multicolumn{2}{c}{\urlprefix{URL prefix}} & & \urlquery{query} & & \urlpage{offset}
    \end{tabular}\\[2.5ex]
    \Nb\textit{Path segment:}\\[1.5ex]
    \begin{tabular}{@{}l@{}c@{}c@{}c@{}c@{}c@{}c@{}c@{}}
        \urlignored{scheme} & \urlignored{network loc.} & \multicolumn{5}{c}{\urlignored{path}} \\
        \arrayrulecolor{urlignored}%
        \cmidrule(r{0.25em}){1-1}%
        \cmidrule(l{0.25em}r{0.25em}){2-2}
        \cmidrule(l{0.25em}){3-7} \\[-2.25ex]
        \texttt{\urlignored{https://}} & \texttt{\urlprefix{chefkoch.de}} & \texttt{\urlprefix{/rs/}} & \texttt{\urlpage{s\textbf{0}}} & \texttt{\urlignored{/}} & \texttt{\urlquery{\textbf{backen}\%20\textbf{dinkelmehl}}} & \texttt{\urlignored{/Rezepte.html}} \\[0.75ex]
        \arrayrulecolor{urlprefix}
        \cmidrule(r{0.35em}){2-3} \\[-4.2ex]
        \arrayrulecolor{urlpage}
        \cmidrule(l{0.15em}){4-4} \\[-4.25ex]
        \arrayrulecolor{urlquery}
        \cmidrule{6-6}
        & \multicolumn{2}{c}{\urlprefix{URL prefix}} & \hspace{-0.5em}\urlpage{page}\hspace{-0.5em} & & \urlquery{query}
    \end{tabular}
\end{minipage}
\caption{Illustration of the components of SERP URLs and the relevant parts for query extraction by \Na query parameter or \Nb path segment.}
\label{figure-url-decomposition}
\end{figure}

To determine the query, page, and offset of the parsers for each search provider, we manually examined the captured URLs in a similar way as for URL prefixes (see Section~\ref{provider-domains-urls}) and derive suited URL parser types and parameters. Regular expressions are optionally used to limit parsing to specific URLs, and to further refine the parsers (e.g., removing prefixes such as \textttsmall{page-} in \textttsmall{/search/page-4}). The resulting set of parsers is applied to all available captures of all search providers, ordered by preference, so that the first parser that returns a non-empty query, page, or offset is used.

Altogether a total of 356.5M~URLs containing queries are collected. Again, the majority of queries stems from search engines (162M~queries,~46\,\%) such as Google (73M~queries,~20\,\%) and Baidu (70M~queries,~20\,\%). On average, 648,092~queries are extracted per search provider. This unfiltered set of queries contains large amounts of duplicates (288M,~81\,\%) for which we identify three reasons:
\Ni
the same query is captured at different times,
\Nii
the query is captured at approximately the same time but with different result page offsets, and
\Niii
the same query is captured as issued from different users (e.g., if a user identifier is included in the URL).
This is supported by the fact that search engines are the main contributors of duplicates~(131M); however, government sites have the highest share~(91\,\%~of all queries from government sites).

We create a set of unique queries for each search provider by selecting a representative query URL (the capture with the shortest query string) from each group with the same parsed query.%
\footnote{As split according to RFC~2396.}
If a group of duplicates has multiple captures with the same query parameter and URL length, the representative URL is chosen by lexicographic order. We refrain from using the capture's timestamp as a tie-breaker to not favor older or newer captures. The deduplication results in 64.5M~unique queries across the final list of 550~search providers. There are 117,353~unique queries per each provider on average. Again, the search engines make up the majority of deduplicated queries~(31M,~45\,\%).

\subsection{SERP Acquisition and Parsing}
\label{serp-acquisition-and-parsing}

\begin{figure}[tb]
  \includegraphics[width=\linewidth]{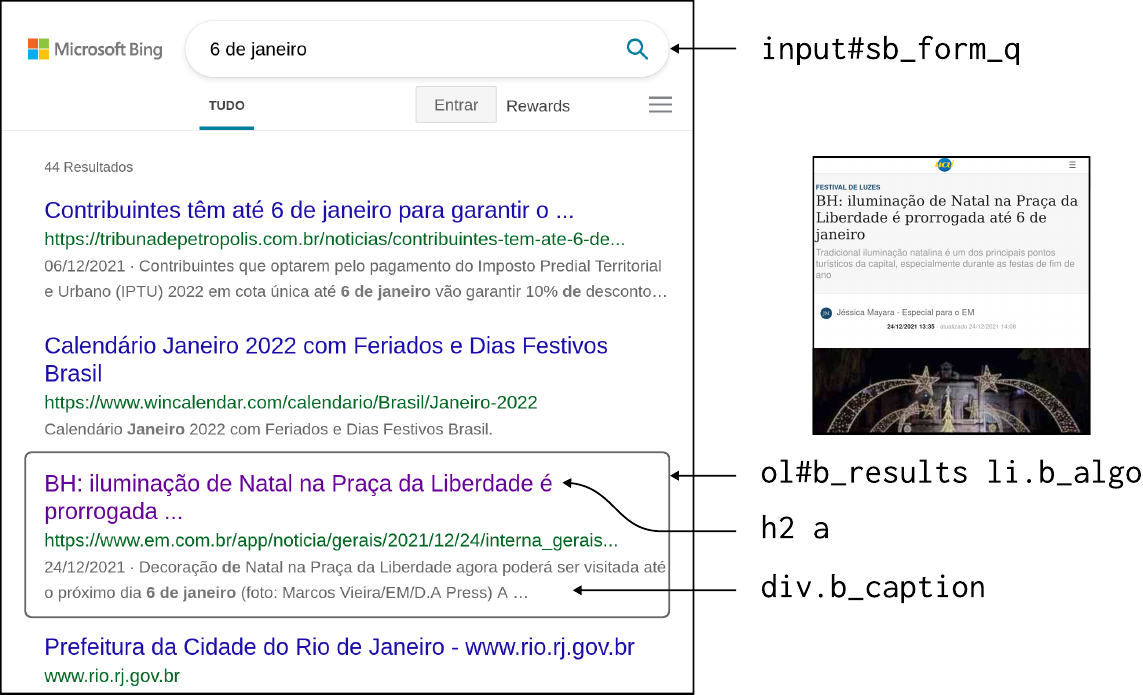}
  \caption{Screenshot of an archived Bing SERP along with the CSS selectors for the query and result items. The nearest archived snapshot of the search result is shown on the right.}
  \label{screenshot-serp-url}
\end{figure}


Previous query logs rarely contain results for the logged queries (see Section~\ref{part2}). However, since the AQL is obtained from web page captures in the Internet Archive, it naturally contains the full ranking of results for many of its queries. By downloading and parsing the archived SERPs, one has access to the full ranking of results for each search query (including the processed query itself, as it appears in the query field of the SERP). Parsing the search results including result titles, referenced URLs, snippets, and the query from a SERP facilitates not only the comparison of different search provider's ranking functions but also the evaluation of their query understanding and reformulation techniques.

We download the SERP HTML content for the unique search queries identified in the previous step and save it in WARC format:%
\footnote{ISO 28500:2017; \url{https://iipc.github.io/warc-specifications/}}
For the 20~most popular search providers SERPs are downloaded for all unique URLs; for the remaining providers the download is limited to a maximum of~100,000 due to resource constraints. Connection timeouts and other errors during download are handled by repeating the download up to 10~times, after which we consider the archived SERP snapshot to be unavailable. Based on the completed downloads, we estimate a total of 137.3M~SERPs to be available, most of which originate from search engines (ca.~46\,\%) and social media platforms (ca.~16\,\%). The downloads are ongoing, and we plan on scaling them up (see Section~\ref{part5}) to compile the full set of estimated SERPs available (see Section~\ref{part4}).

From the downloaded SERPs the search result ranking is extracted and processed using a configurable parser pipeline based on FastWARC~\cite{bevendorff:2021b}, Beautiful Soup,%
\footnote{\url{https://pypi.org/project/beautifulsoup4/}}
and Approval Tests.%
\footnote{\url{https://pypi.org/project/approvaltests/}}
In detail, the CSS path or selector%
\footnote{See \url{https://facelessuser.github.io/soupsieve/} for a list of supported selectors. The CSS path to an HTML element can be inferred using a web browser's developer tools.}
to the result list items is specified, as well as the path from each individual result item to its title element, the referenced URL anchor, and the snippet text. Similarly, the processed query is parsed based on the CSS path to the query field. Figure~\ref{screenshot-serp-url} shows an archived Bing SERP and highlights how CSS paths are used to select relevant HTML tags. Each search provider can have multiple parser configurations, ordered by preference, that, for example, account for a changed HTML structure after redesigns of a search provider's SERPs.

We derive parser configurations (CSS paths) for the 50~most popular search providers by generating Approval Tests according to following workflow for a provider:
\Ni
Randomly sample 10~SERPs from the downloaded SERPs.
\Nii
For each SERP manually annotate the expected ranking and query.
\Niii
Apply the existing parser configurations to the sampled SERPs.
\Niv
Compare the parsed results to the annotations.
\Nv
If the annotations do not match, inspect the HTML page in a browser, adapt or extend missing patterns, and add them to the provider's parser configurations.
New configurations are added iteratively until all sampled SERPs are correctly parsed. Altogether, 70~parser configurations for SERPs and 57~parsers for processed queries of the 50~most popular providers are derived. With additional manual tests for the 10~most popular providers (see Table~\ref{table-archive-query-log-top-10}), the parsers pass a test suite of 444~Approval Tests. The code base is available open source.%
\footnote{AQL code: \url{https://github.com/webis-de/archive-query-log}}

\subsection{The Archive Query Log~2022 (AQL-22)}
\label{aql-22}

We merge the filtered URL captures, queries, and SERPs into a single corpus to be used in subsequent analyses (see Section~\ref{part4}). This corpus, the Archive Query Log~2022, consists of two artifacts:
\Ni
a set of queries and
\Nii
a set of ranked documents (search result snippets).
Both artifacts are stored in a GZIP-compressed, newline-delimited JSON format.%
\footnote{\url{https://jsonlines.org/}}
To create the query set, each captured URL is assigned a unique identifier based on the full URL string and timestamp of the capture.%
\footnote{Name-based SHA-1 UUID according to RFC~4122: \url{https://rfc-editor.org/rfc/rfc4122}}
The captured URL is associated with its parsed query, the location of the stored copy of the SERP, and the processed query and search results parsed from that SERP. In addition, we include the URL to the SERP's archived snapshot on the Wayback Machine and tag the query language based on the parsed query text using \cld.%
\footnote{Google's Compact Language Detector; \url{https://github.com/google/cld3}}

The set of result documents is created by concatenating all ranked search results (i.e., rank, snippet text with title, and document URL to the referenced web page) from all parsed SERPs. Each document is assigned a unique identifier based on the document URL, the timestamp of its origin query, and the rank of the snippet on the SERP. We also associate each document with the attributes of the corresponding query in the query set. Two additional fields are the URL to the nearest available snapshot of the SERP on the Wayback Machine and the snippet language as tagged using \cld based on the snippet's title and text.

\begin{table*}[ht!]
\small
\centering
\caption{The Archive Query Log~2022~(AQL-22) in detail. Categories manually annotated. Top-3 languages tagged by \cld. Ticks in the timelines indicate days of archival from Jan~1999 to Dec~2022. Number of SERPs and results estimated (c.f.\ Section~\ref{estimates}).}
\label{table-archive-query-log-detail}
\newcommand{\providerlogo}[1]{\raisebox{-0.25ex}{\includegraphics[height=2ex]{logo-#1}}}
\newcommand{\providertimeline}[1]{\includegraphics[height=1.25ex,width=5em,trim=7.5pt 7.5pt 7.5pt 7.5pt,clip=true]{timelines/#1}}
\renewcommand{\arraystretch}{0.9}
\renewcommand{\tabcolsep}{3pt}
\vspace*{-3pt}
\begin{tabular*}{\linewidth}{@{}lll@{\extracolsep{\fill}}rrrlcrrlc@{}}
\toprule
\multicolumn{2}{@{}l}{\textbf{Search provider}} & \multicolumn{1}{l}{\textbf{Category}} & \multicolumn{1}{c}{\textbf{URLs}} & \multicolumn{4}{c}{\textbf{Queries}} & \multicolumn{1}{c}{\textbf{SERPs}} & \multicolumn{3}{c}{\textbf{Results}} \\
\cmidrule(l@{\tabcolsep}r@{\tabcolsep}){4-4}
\cmidrule(l@{\tabcolsep}r@{\tabcolsep}){5-8}
\cmidrule(l@{\tabcolsep}r@{\tabcolsep}){9-9}
\cmidrule(l@{\tabcolsep}){10-12}
& & &\multicolumn{1}{c}{Total} & \multicolumn{1}{c}{Total} & \multicolumn{1}{c}{Unique} & \multicolumn{1}{c}{Lang.} & \multicolumn{1}{c}{Timeline} & \multicolumn{1}{c}{Estimate} & \multicolumn{1}{c}{Estimate} & \multicolumn{1}{c}{Lang.} & \multicolumn{1}{c@{}}{Timeline} \\
\midrule
\providerlogo{google}             & Google        & Search engine  &    89,364,948 &  72,673,044 & 19,953,592 & en, th, zh &   \providertimeline{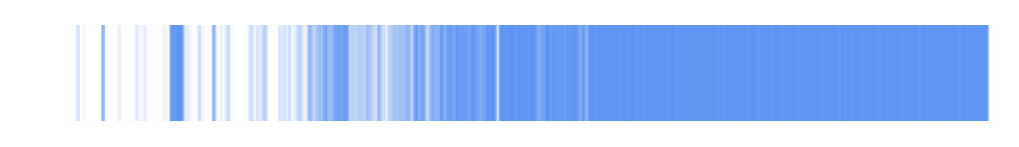}   &  27,979,122 &   223,073,409 & en, de, pt &   \providertimeline{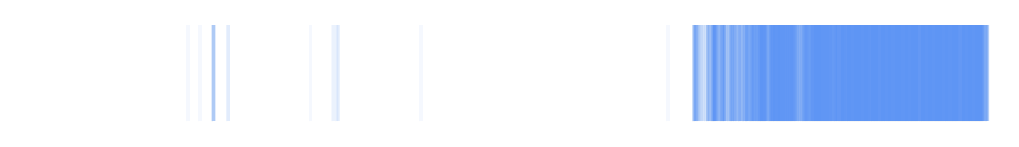}   \\
\providerlogo{youtube}            & YouTube       & Media sharing  &    41,846,525 &  41,365,166 & 11,250,179 & ru, ko, ja &  \providertimeline{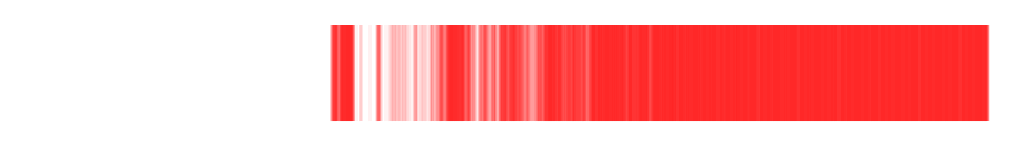}   &  15,925,589 &   339,164,817 & ru, en, ko &  \providertimeline{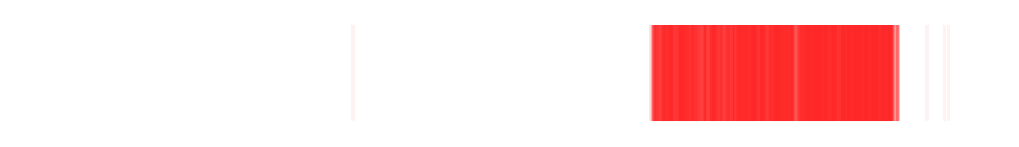}   \\
\providerlogo{baidu}              & Baidu         & Search engine  &    78,506,825 &  69,619,339 &  2,900,878 & zh, ga, ja &   \providertimeline{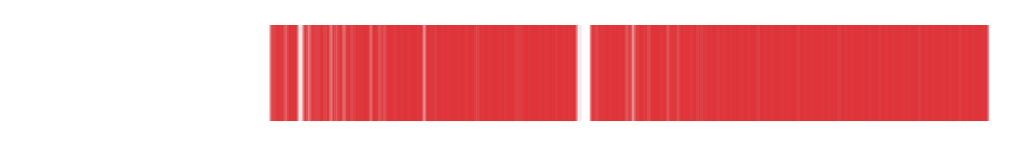}    &  26,803,445 &   107,623,054 & zh, en, mr &   \providertimeline{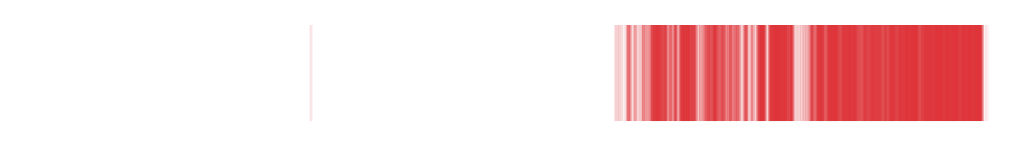}    \\
\providerlogo{qq}                 & QQ            & Web portal     &       515,895 &     513,608 &     51,228 & zh, ja, lb &     \providertimeline{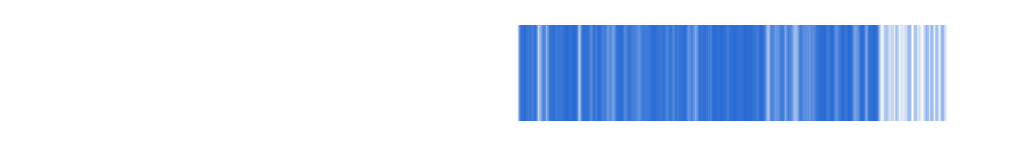}     &     197,739 &     2,101,929 & --         &     \providertimeline{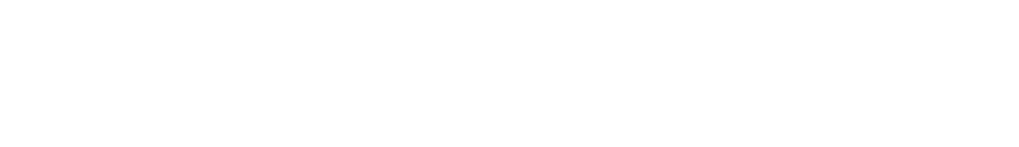}     \\
\providerlogo{facebook}           & Facebook      & Social media   &     3,131,212 &     159,087 &     35,492 & ca, en, bs &  \providertimeline{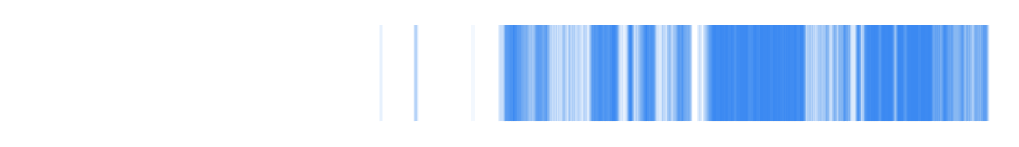}  &      61,249 &       651,065 & --         &  \providertimeline{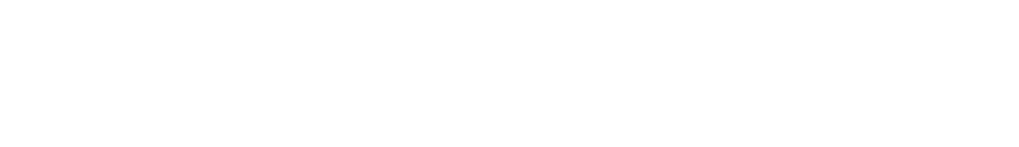}  \\
\providerlogo{yahoo}              & Yahoo!        & Web portal     &     8,787,707 &   2,827,103 &  1,232,589 & en, la, de &   \providertimeline{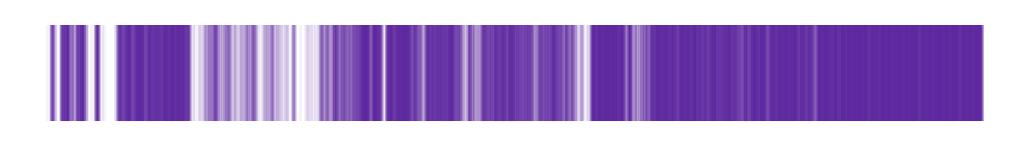}    &   1,088,435 &     9,242,960 & en, es, pt &   \providertimeline{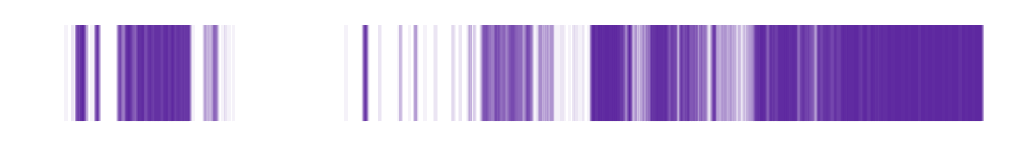}    \\
\providerlogo{amazon}             & Amazon        & E-commerce     &    66,795,164 &     776,127 &    315,068 & en, ja, zh &   \providertimeline{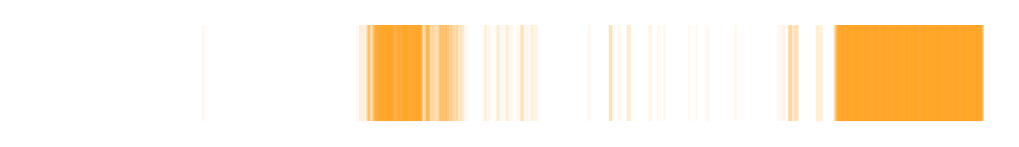}   &     298,809 &     7,789,151 & en, ja, it &   \providertimeline{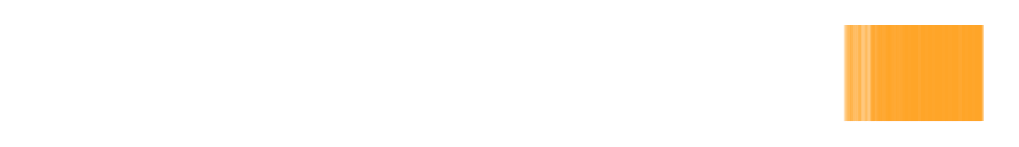}   \\
\providerlogo{wikipedia}          & Wikipedia     & Wiki           &    68,547,509 &   1,707,058 &    621,971 & sv, zh, en & \providertimeline{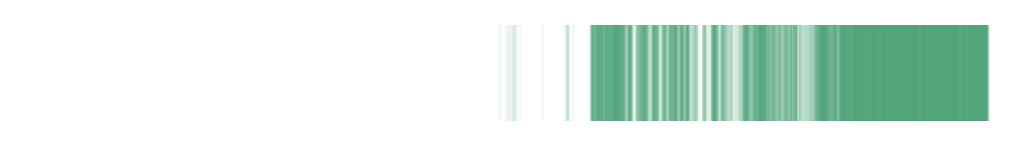}  &     657,218 &     6,986,104 & --         & \providertimeline{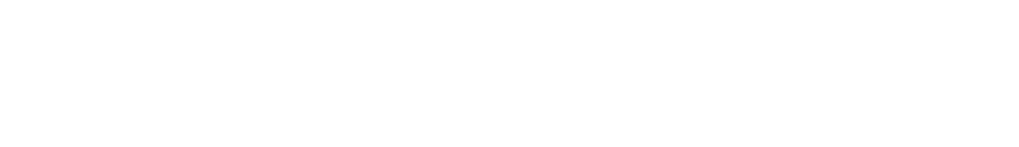}  \\
\providerlogo{jd-dot-com}         & JD.com        & E-commerce     &     4,370,884 &   3,902,604 &    370,473 & zh, hr, ja &     \providertimeline{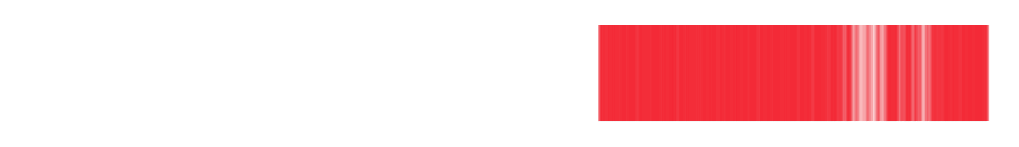}     &   1,502,503 &    15,971,325 & --         &     \providertimeline{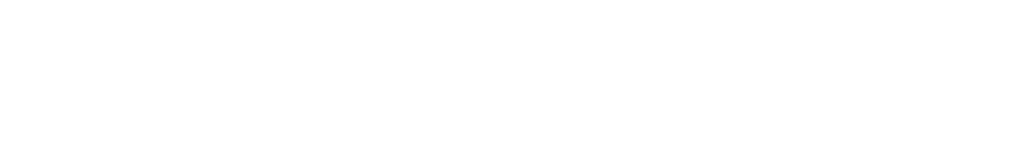}     \\
\providerlogo{360}                & 360           & Search engine  &     1,495,365 &   1,090,152 &     65,596 & zh, ja, mg &    \providertimeline{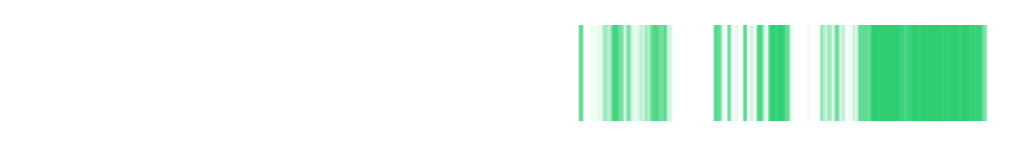}     &     419,708 &     3,487,458 & zh, mr, en &    \providertimeline{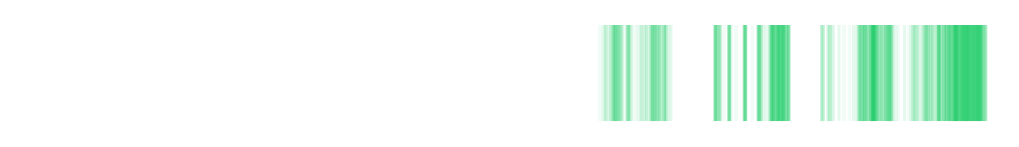}     \\
\providerlogo{weibo}              & Weibo         & Social media   &     6,245,012 &   5,324,385 &  1,886,458 & zh, ja, en &   \providertimeline{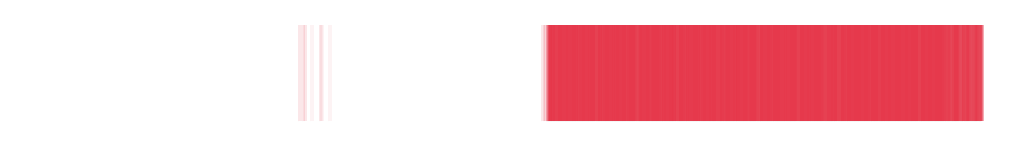}    &   2,049,889 &    21,789,936 & --         &   \providertimeline{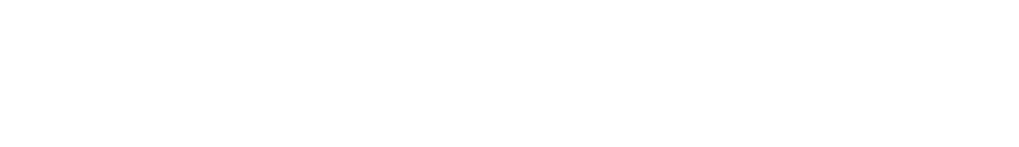}    \\
\providerlogo{reddit}             & Reddit        & Forum          &        94,162 &      89,492 &     36,852 & en, la, de &   \providertimeline{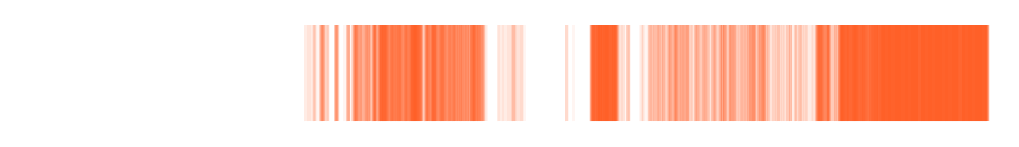}   &      40,369 &       429,115 & --         &   \providertimeline{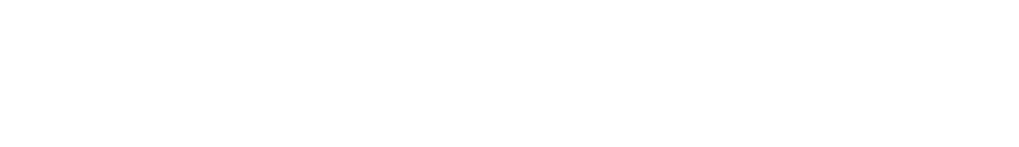}   \\
\providerlogo{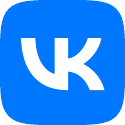}         & Vk.com        & Social media   &       643,354 &     153,642 &     46,134 & ru, sr, ky &     \providertimeline{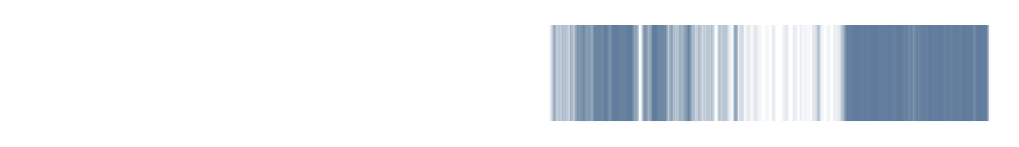}     &      59,152 &       628,775 & --         &     \providertimeline{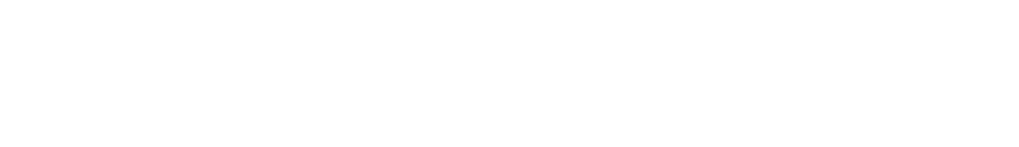}     \\
\providerlogo{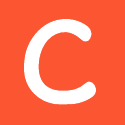}               & CSDN          & Social media   &        21,863 &         946 &        736 & zh, en, vi &    \providertimeline{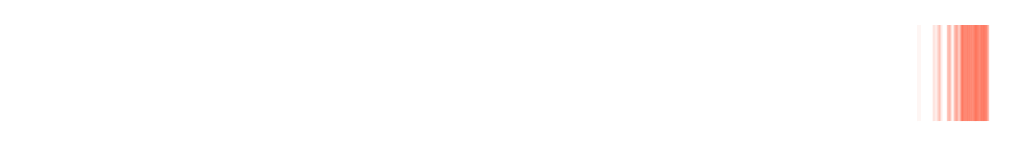}    &         810 &         8,610 & --         &    \providertimeline{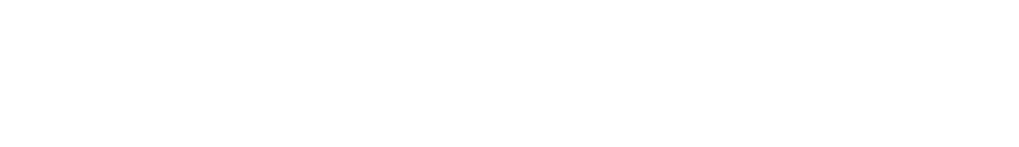}    \\
\providerlogo{bing}               & Bing          & Search engine  &    11,263,539 &   6,152,425 &  2,253,965 & en, zh, pt &    \providertimeline{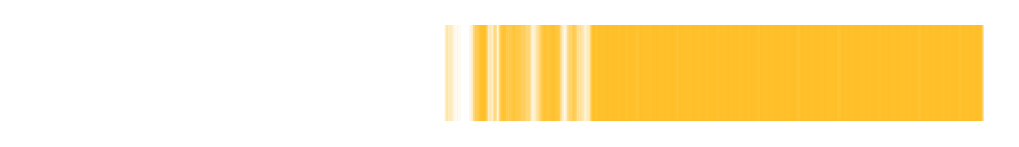}    &   2,368,684 &    12,625,017 & en, pt, fr &    \providertimeline{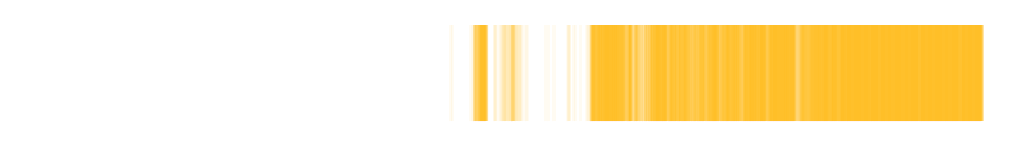}    \\
\providerlogo{twitter}            & Twitter       & Social media   &    55,499,532 &  48,084,528 &  3,869,382 & ja, en, gl &  \providertimeline{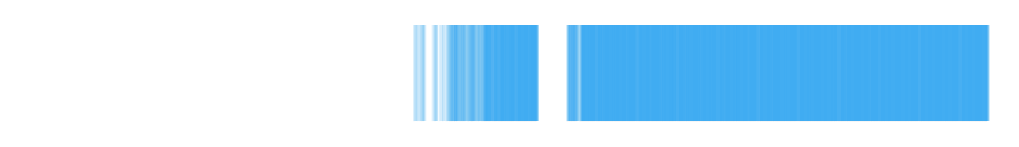}   &  18,512,544 &   241,835,785 & en, ja, es &  \providertimeline{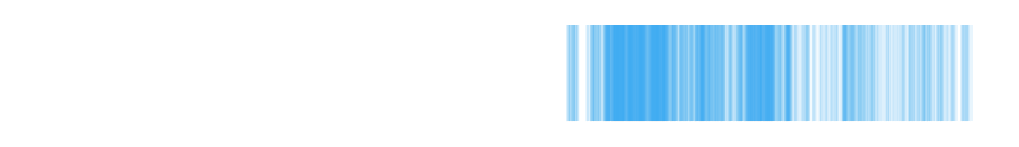}   \\
\providerlogo{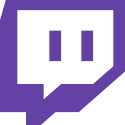}             & Twitch        & Streaming      &        21,931 &      15,225 &     11,445 & en, zh, de &   \providertimeline{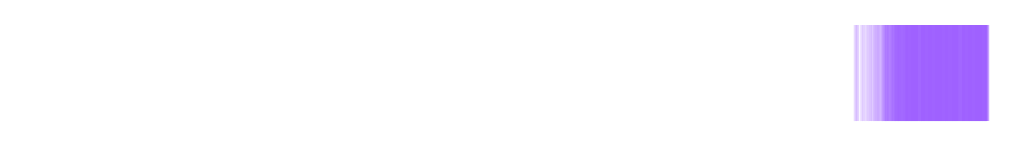}   &      12,587 &       133,797 & --         &   \providertimeline{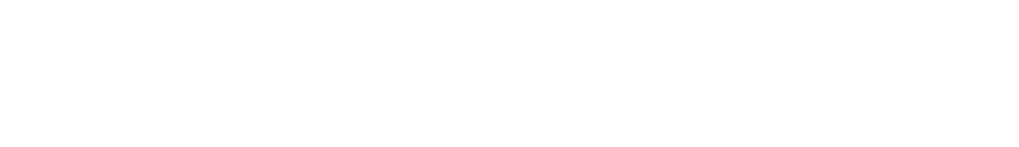}   \\
\providerlogo{ebay}               & eBay          & E-commerce     &     7,927,123 &   5,507,532 &  1,379,646 & zh, en, la &    \providertimeline{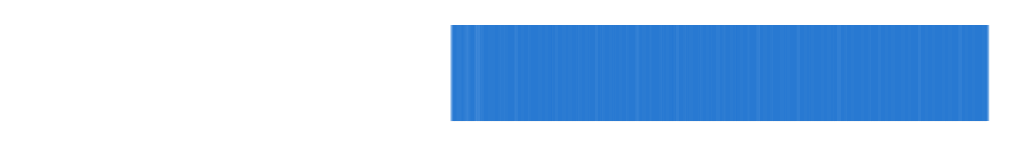}    &   2,120,400 &    25,130,312 & en, es, de &    \providertimeline{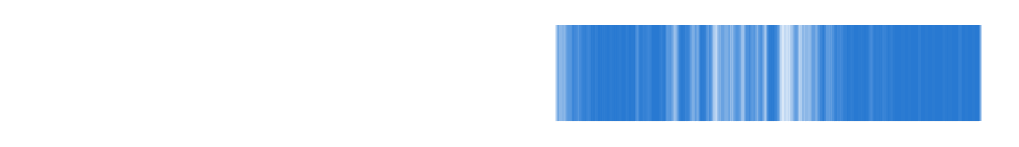}    \\
\providerlogo{naver}              & Naver         & Search engine  &     1,063,991 &     756,153 &    400,490 & ja, ko, vi &   \providertimeline{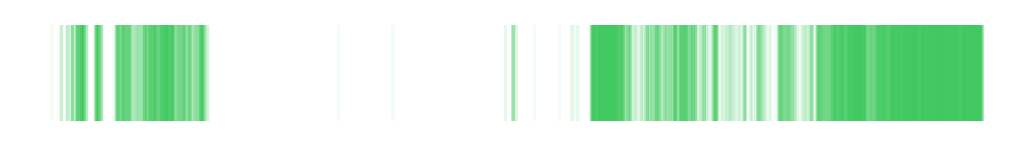}    &     291,119 &     2,938,902 & ko, en, hi &   \providertimeline{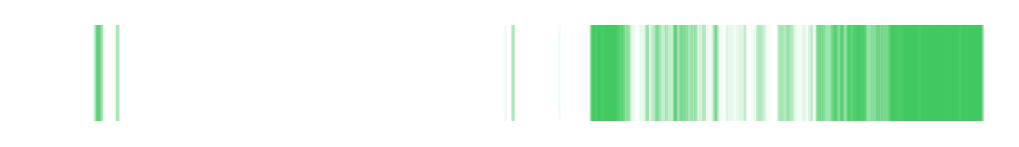}    \\
\providerlogo{aliexpress}         & AliExpress    & E-commerce     &     4,620,331 &   1,861,642 &     55,849 & en, lb, fy & \providertimeline{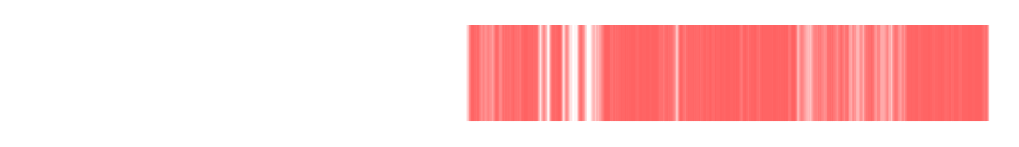} &     716,732 &     5,719,031 & en, fr, ru & \providertimeline{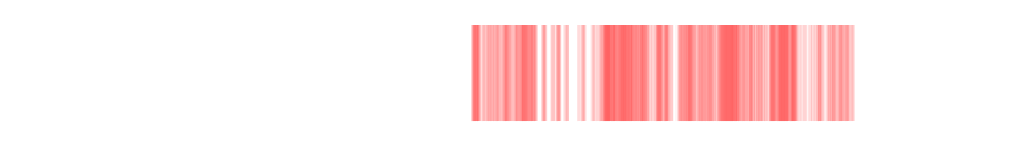} \\
\midrule
\hspace{0.4em}\rotatebox{90}{...} & 530~others    &                &   559,225,614 &  93,871,236 & 17,806,322 & en, zh, de & \providertimeline{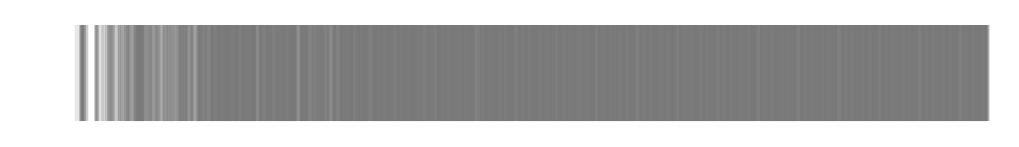}  &  36,194,558 &   382,644,117 & en, zh, de & \providertimeline{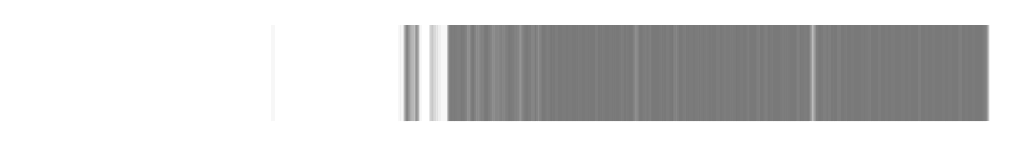}  \\
\midrule
\(\sum\)                          & 550           &                & 1,009,988,486 & 356,450,494 & 64,544,345 & zh, en, ga &    \providertimeline{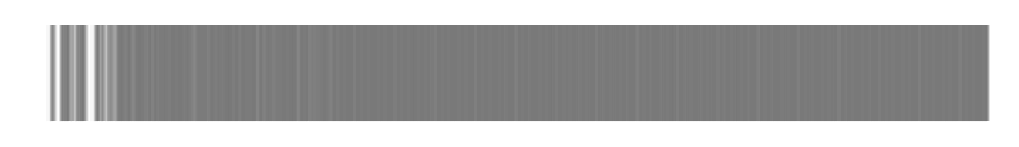}     & 137,300,661 & 1,409,974,669 & en, ru, ko &    \providertimeline{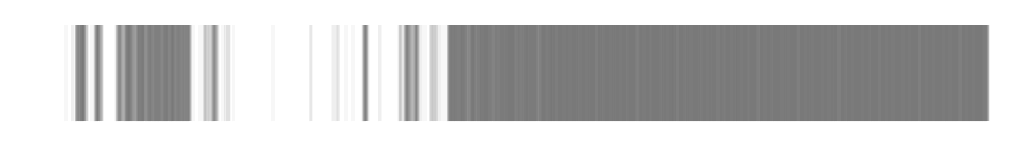}     \\
\bottomrule
\end{tabular*}
\end{table*}

\section{Analysis}
\label{part4}

In order to provide a better understanding of the corpus, we conduct some analyses on the query and SERP characteristics and highlight potential use cases. Detailed analyses of the AQL will be the subject of future work. At the time of writing, we could download and parse all URLs and queries. However, due to computational constraints, only a subset of all available SERPs could be parsed yet (see Section~\ref{estimates}) and are used for our analyses.

\subsection{Query Characteristics}

\begin{figure}[tb]
    \includegraphics[width=1\linewidth]{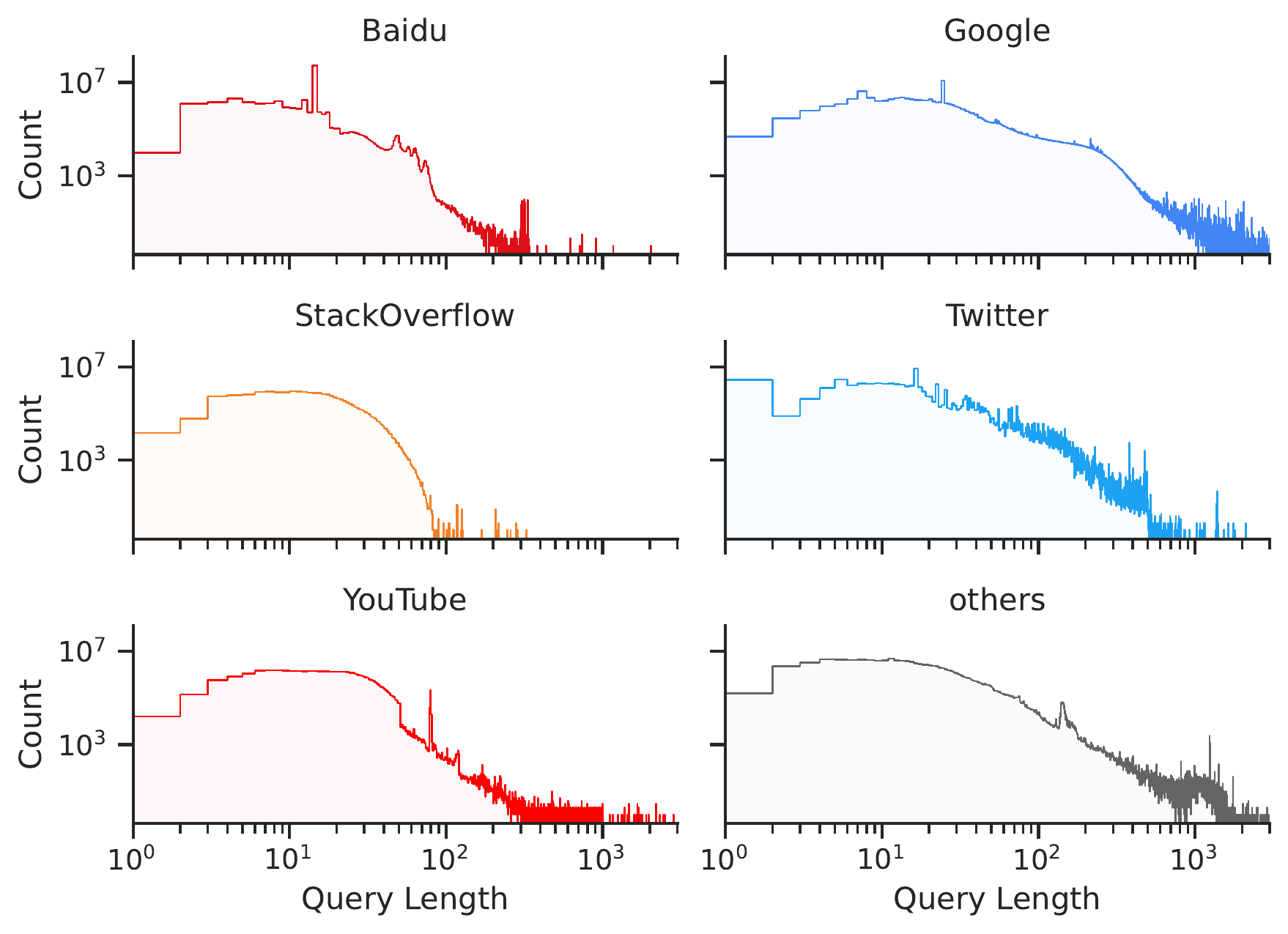}
    \vskip-2ex
    \caption{Distribution of query lengths for 5~search providers contributing the highest amount of queries. The remaining search providers are grouped as ``others''.}
    \label{figure-query-length}
\end{figure}

A central feature of the AQL is its diversity. In addition to the variety of search providers, it also features a total of 104~different languages%
\footnote{Out of 107~detectable with \texttt{cld3}.}
with Chinese and English being the most frequently used query languages (see Table~\ref{table-archive-query-log-detail}). The query length in the AQL follows a skewed distribution, with most queries containing between~5 and 20~characters. Figure~\ref{figure-query-length} provides a visualization for the 5~search providers with the most queries. We inspect samples of queries with 5, 10, 100, and 1000~characters. Very short queries are often Mandarin keywords (e.g, \raisebox{-0.1ex}{\includegraphics[height=1.5ex]{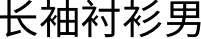}}, ``men's long sleeve shirt') or single English words (e.g., \query{video}). Queries with 10~characters are mostly keyword-style queries from Latin languages (e.g., \query{comic font}) or hashtags (e.g. \raisebox{-0.05ex}{\includegraphics[height=1.3ex]{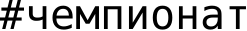}}, ``championship''). Most longer queries extensively use search operators like \textttsmall{site:} and \textttsmall{order:}, include literature references, or include long multi-line text like stack traces from errors in programming.

\begin{figure}[tb]
    \includegraphics[width=\linewidth]{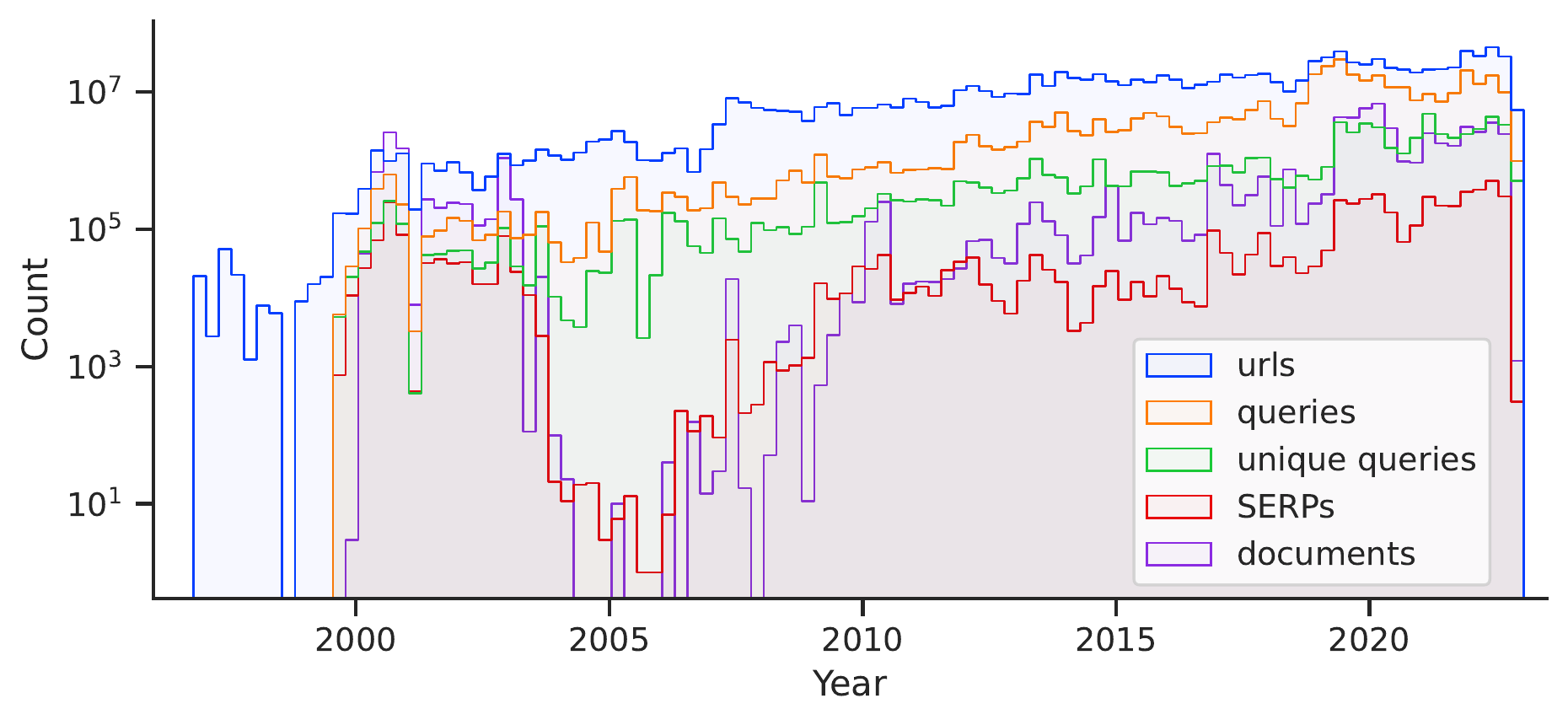}
    \vskip-2ex
    \caption{Time coverage of the total amount of different data types collected for the AQL-22, per quarter.}
    \label{figure-time-coverage}
\end{figure}

Second, we evaluate whether obscene or unwanted terms comprise a large share of the AQL. We use lists of obscene words for 27~languages%
\footnote{Compiled by Shutterstock; \url{https://github.com/LDNOOBW/List-of-Dirty-Naughty-Obscene-and-Otherwise-Bad-Words}.}
and expand the list of English terms with new expressions found in the downloaded queries. We check each query from the two most frequent languages, Chinese and English, for their lists of obscene terms. Overall, only 1.30\,\%~of all queries contain obscene terms. The highest share of these obscene queries were observed on pornography~(19.48\,\%), torrent~(3.73\,\%), and forum~(2.87\,\%) websites. For non-pornographic search providers, most stem from \domain{heroturko.org}~(16.13\,\%, e-commerce), \domain{reddit.com}~(5.05\,\%, forum), and \domain{kat.cr}~(4.08\,\%, forum).

Regarding time coverage, Figure~\ref{figure-time-coverage} shows that the archival of SERPs dropped between~2004 and~2010 for unknown reasons, which might indicate that using more specialized SERP parsers are required, or that results were loaded using Javascript. The queries, however, extend over the whole timespan, with tens of thousands of queries recorded for the early~2000's as well. Table~\ref{table-archive-query-log-detail} contains an overview of the 20~most popular services' time coverage.

\subsection{SERP Characteristics}

\begin{table}[t!]
\caption{Most frequent document domains in the top-5 or the top-10 search results compared to references to the search provider's own domain~(\raisebox{-0.2ex}{\(\circlearrowleft\)}) or 791,646~other domains~(\raisebox{0.5ex}{...}). }
\label{table-referenced-domains}
\small
\renewcommand{\tabcolsep}{3.0pt}
\renewcommand{\arraystretch}{0.85}
\begin{tabular}{@{}*{12}{c}@{}}
\toprule
\textbf{Top}
& \raisebox{-0.5ex}{\includegraphics[height=2ex]{logo-wikipedia}} 
& \raisebox{-0.5ex}{\includegraphics[height=2ex]{logo-youtube}} 
& \raisebox{-0.5ex}{\includegraphics[height=2ex]{logo-facebook}}
& \raisebox{-0.5ex}{\includegraphics[height=2ex]{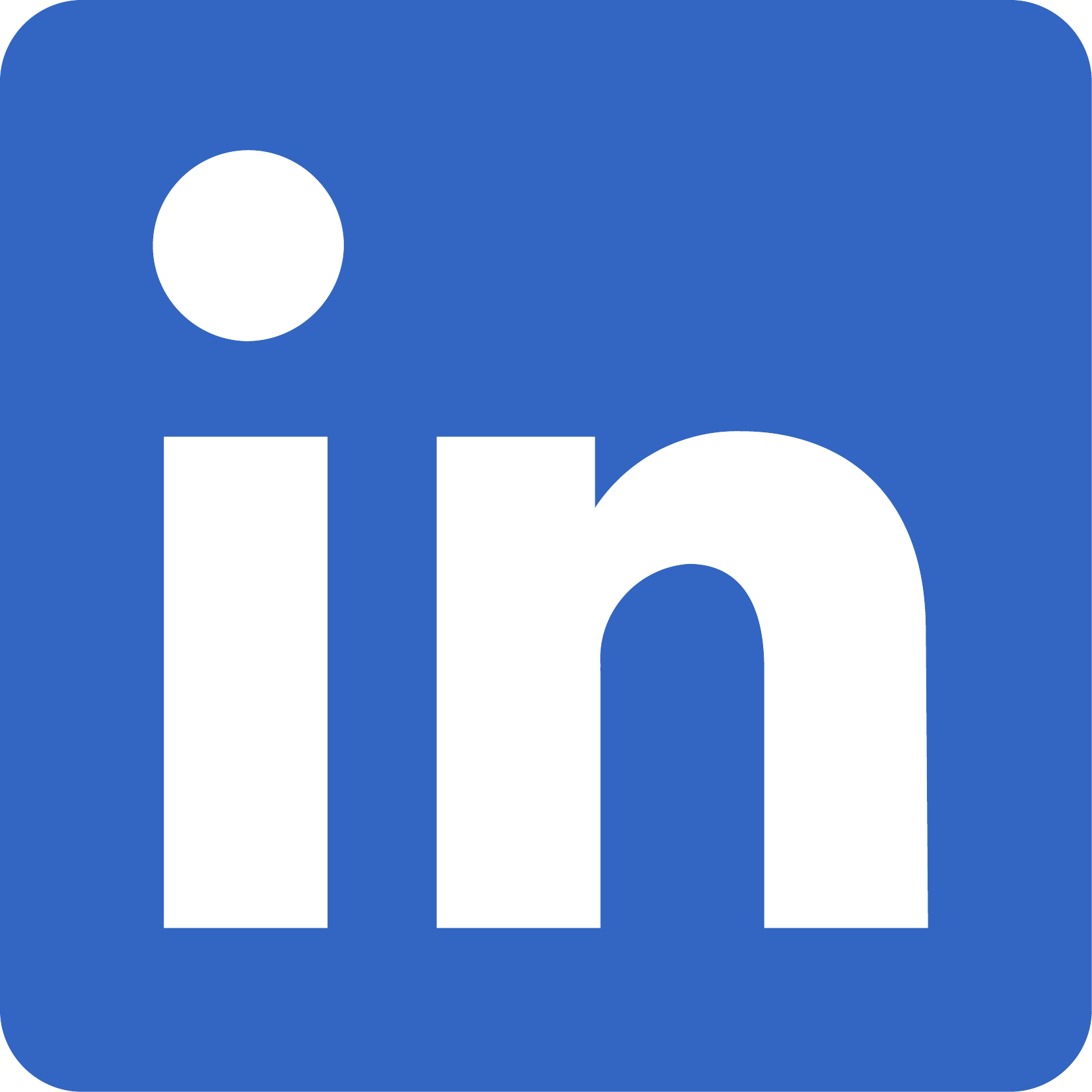}}
& \raisebox{-0.5ex}{\includegraphics[height=2ex]{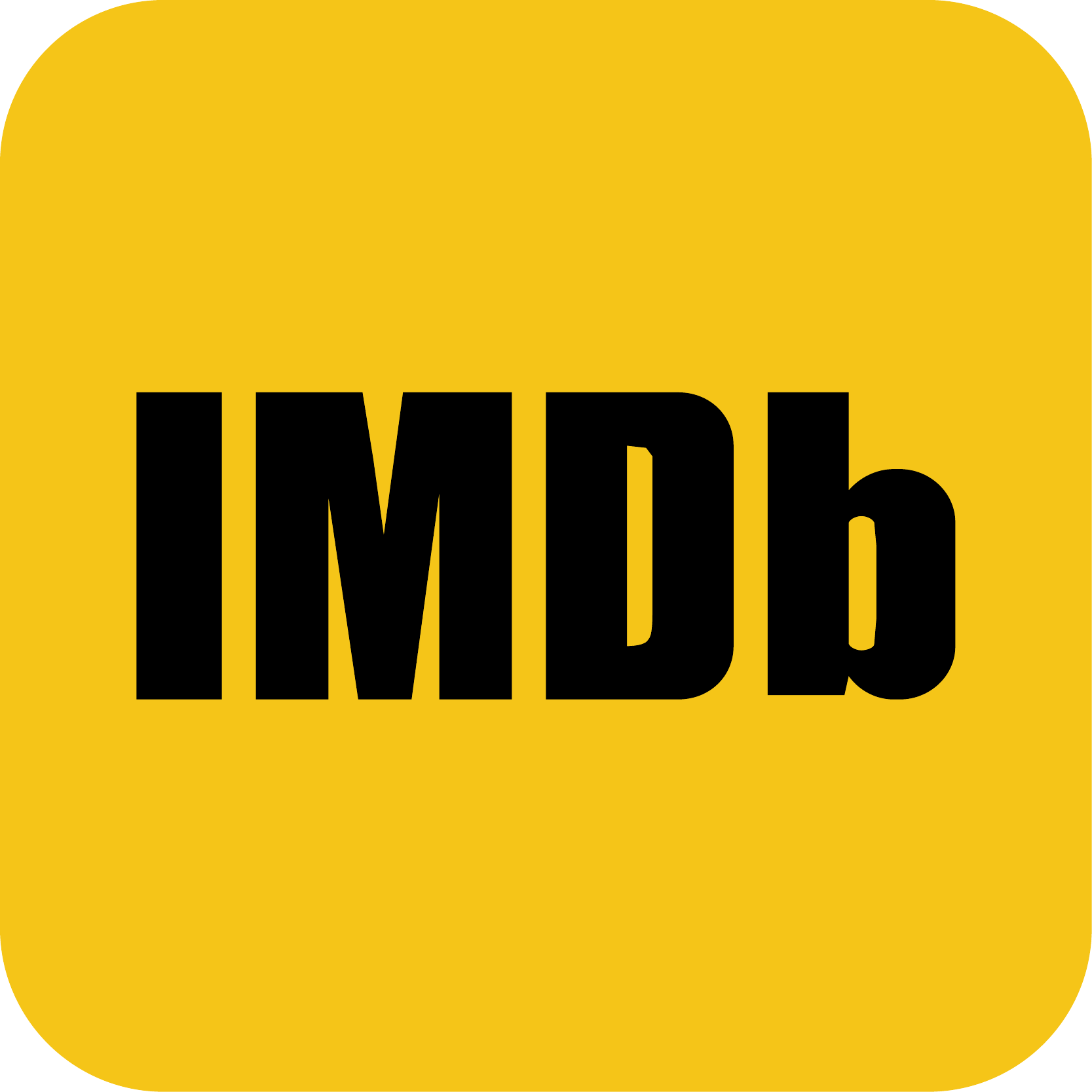}}
& \raisebox{-0.5ex}{\includegraphics[height=2ex]{logo-instagram}} 
& \raisebox{-0.5ex}{\includegraphics[height=2ex]{logo-amazon}}
& \raisebox{-0.5ex}{\includegraphics[height=2ex]{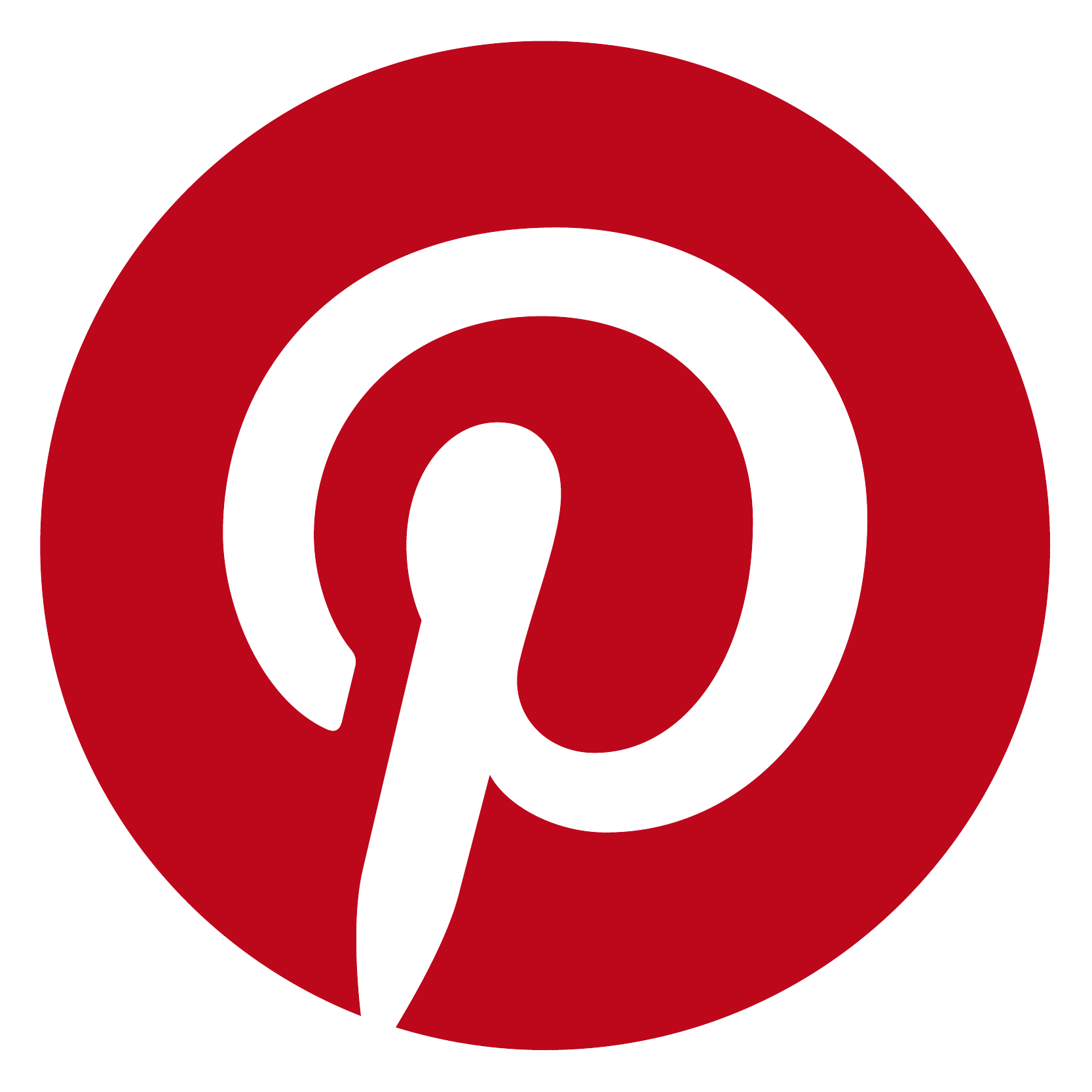}} 
& \raisebox{0.5ex}{...}
& \raisebox{-0.2ex}{\(\circlearrowleft\)}
\\
\midrule
5  & 2.9\,\% & 0.8\,\% & 0.6\,\% & 0.4\,\% & 0.3\,\% & 0.3\,\% & 0.3\,\% & 0.2\,\% & 24.3\,\% & 69.6\,\% \\
10 & 2.2\,\% & 0.7\,\% & 0.5\,\% & 0.3\,\% & 0.3\,\% & 0.2\,\% & 0.3\,\% & 0.3\,\% & 24.6\,\% & 70.4\,\% \\
\bottomrule
\end{tabular}
\end{table}

In Table~\ref{table-referenced-domains}, we consider the most frequently referenced domains from search result URLs as an indicator of plausible rankings. Excluding frequent self-references~(e.g., to internal redirect pages), by far the most frequently ranked domain is \domain{wikipedia.org} contributing 2.9\,\%~of all top-5 results and 2.2\,\%~of the top-10. Other popular domains like \domain{youtube.com} and \domain{facebook.com} also frequently appear on high ranks. The most frequent languages are shown in Table~\ref{table-archive-query-log-detail}. Interestingly, Chinese is not among the top-3 languages of search results, even though it is the most frequently used query language, representing a bias that should be evaluated in future work.

\subsection{Use Cases}

The AQL opens up a variety of use cases for the IR community. We highlight two promising applications.

\begin{figure}
    \includegraphics[width=\linewidth]{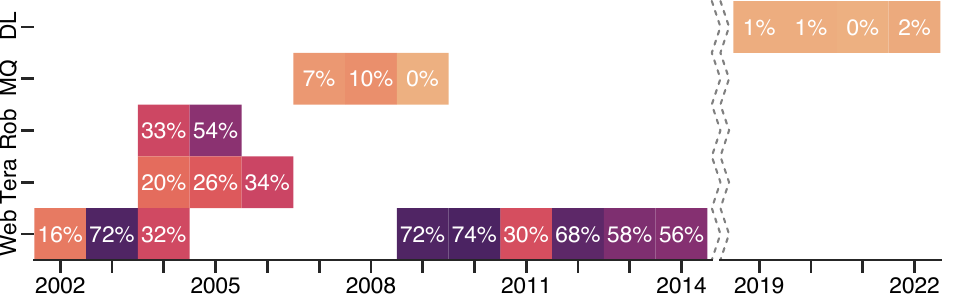}
    \caption{Query overlap with TREC Robust~(Rob), Terabyte (Tera), Million Query~(MQ), Web, and Deep Learning~(DL).}
    \vspace{4pt}
    \label{figure-trec-overlap}
\end{figure}

First, we evaluate the exact overlap of the queries in the AQL with the collections used in various TREC tracks from~2004 to~2022~\cite{voorhees:2004,voorhees:2005,clarke:2004,clarke:2005,allen:2007,allen:2008,carterette:2009,craswell:2002,craswell:2003,craswell:2004,clarke:2009,clarke:2010,clarke:2011,clarke:2012,thompson:2013,thompson:2014,craswell:2019,craswell:2020b}. As shown in Figure~\ref{figure-trec-overlap}, the highest overlap exists with the Web tracks, specifically in~2010~(74\,\%), 2003, and~2009~(both~72\,\%). The lowest overlap was found with the Deep Learning tracks, ranging between~0--2\,\%. The high overlap on older Web tracks poses an interesting opportunity for enriching existing benchmarks. While query logs have been used previously in shared community tasks~\cite{mandl:2009}, shared tasks often specify only one query for each topic. We propose to sample semantically similar queries from the AQL to generate topics with user query variations~\cite{bailey:2016} automatically. On the other hand, the low overlap with the Deep Learning tracks highlights a sampling bias in creating the Deep Learning topics. The topics were sampled from the official eval set of MS~MARCO, which includes only natural language questions from a Bing query log~\cite{craswell:2019,nguyen:2016}. The AQL, on the other hand, contains a much broader range of queries, including queries from other search providers and non-question-like queries. Therefore, we propose using the AQL to create new, ``harder'' Deep Learning topics that are more representative of other kinds of queries users submit.

\begin{figure}
    \includegraphics[width=\linewidth]{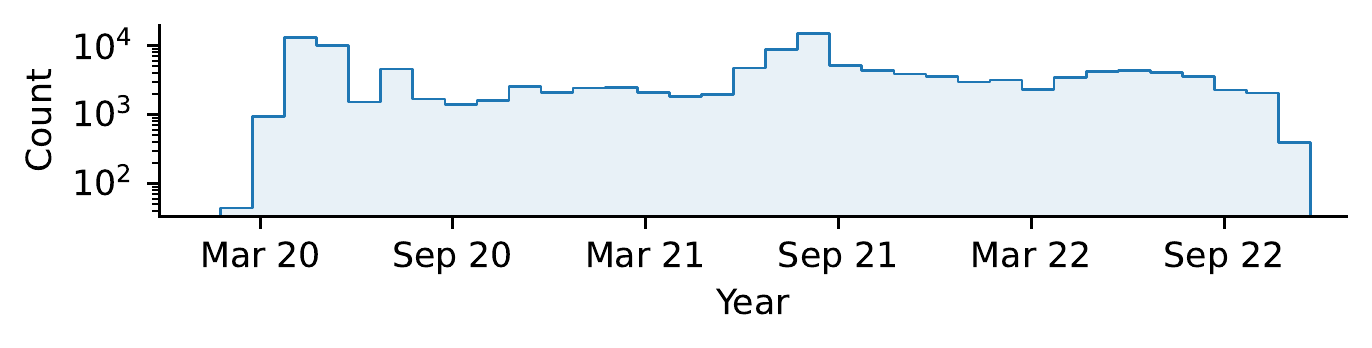}
    \vskip-2ex
    \caption{Timeline of Covid-19-related terms in the AQL-22.}
    \label{figure-time-series-covid}
\end{figure}

Second, we demonstrate how global trends are reflected in the AQL on the example of the Covid-19 pandemic. In Figure~\ref{figure-time-series-covid}, we count the occurrences of the terms \query{covid 19}, \query{sars cov 2}, and \query{corona virus} each month since the outbreak in 2019. A peak can be observed during the first global lockdowns in early~2020, but overall, interest in the pandemic has yet to stagnate. The example showcases how the AQL enables unique opportunities for diachronically analysing global trends.

\subsection{Total Size Estimates}
\label{estimates}

As Section~\ref{part3} explains, we have only downloaded and parsed a subset of all available SERPs. Based on our results so far, we estimate 70\,\%~of all SERP snapshots to be available for download. Assuming an estimated parsing success rate of~55\,\% and 10.6~results per SERP, we expect the total number of parsed SERPs in the AQL-22 to be~137.3M with 1.4B~search results. As outlined in Section~\ref{part5}, we continue to download and parse SERPs and look forward to expanding the AQL with the IR community.

\section{Discussion}
\label{part5}

Access to query logs has long been an insurmountable barrier to answering critical questions about the search economy at large---if not to ask them in the first place. As a consequence, the media and public were left with no choice but to trust search engines on questions such as ``How accountable are search providers in terms of measures of interest, like representation and fairness?'', ``How have these accountability measures changed for these organizations over time?'', and ``How honest has self-reported accountability of these organizations been?''. As the most extensive public query log to date, the AQL enables detailed analyses of and thus facilitates the public discourse on the search industry. It also furthers the research on information retrieval, whose retrieval models are often presumed to be behind or at least detached from those of industry players~(e.g.,~\citet{azzopardi:2016}). Using the AQL, researchers will also be able to answer questions such as ``How far is academic information retrieval research behind industry?'', ``How much do query logs contribute compared to other ranking signals?'', and ``What are domain-specific differences?''.

However, with the scale of the AQL, several ethical and legal considerations also arise, particularly around personally identifiable information or illegal content. To address inherent risks, we release the data by imposing a barrier to access that minimizes potential harms while giving the information retrieval community as much freedom to conduct their research as possible. In addition, we also acknowledge the challenges of creating such extensive collections and discuss our plans for opening contributions to the AQL.

\subsection{Accessing the AQL}
\label{access}

Our goal with releasing the AQL is to do so responsibly and in a privacy-preserving manner. We work towards that goal using the TIRA platform~\cite{froebe:2023b,potthast:2019p} for any analysis on the AQL one wishes to conduct. TIRA has been used since~2012~\cite{gollub:2012c} to facilitate shared tasks with software submissions while ensuring that the submitted software can process the data without giving the participants themselves access. TIRA achieves this through sandboxing, i.e., disconnecting the software from the internet while it is running and thus ensuring that it can not leak data. We added the AQL to TIRA to allow researchers to submit their analysis software as Docker images. The platform is open to the public, and we provide examples and documentation on how to perform analyses on the AQL.%
\footnote{\url{https://tira.io/task/archive-query-log}}
Specific shared tasks will be developed as well.

TIRA allows running arbitrary software packaged in Docker images on the AQL dataset in a privacy-preserving way, as the analysis results are blinded until reviewed. Specifically, we review the output and the software installed in the Docker image to ensure no sensitive data is leaked. TIRA runs the software in a Kubernetes cluster (1,620~CPU cores, 25.4\,TB~RAM, 24~GeForce GTX~1080 GPUs) with a timeout of 24~hours, so that almost any evaluation is supported.%
\footnote{We can extend the timeouts and available resources individually if the need arises.}
In summary, TIRA provides the ideal means to work with the AQL, ensuring sensitive query log data remains secure and is used for academic research in a responsible fashion.

\subsection{Limitations and Scalability}

While creating the AQL, we encountered several technical limitations that guide future optimizations and improvements. First, the various parsers for creating the AQL-22 corpus were written semi-automatically. This approach was error-prone and inefficient, requiring much manual work. When building future versions of the AQL, we plan to train token classification models like BERT~\cite{devlin:2019} to automatically generate query parsers based on a training set derived from our existing parsers. Manually finding the correct CSS paths for snippets on a SERP is a similarly tedious process that can benefit from wrapper generation~\cite{kushmerick:1997}, which has successfully been applied to web page parsing~\cite{chang:2006}. 

Second, dynamic content cannot be interpreted by our existing parsers. The SERPs of DuckDuckGo, for instance, are loaded dynamically using JavaScript and thus cannot be parsed from just the archived HTML snapshot, yet the search results are still archived as a different record. To overcome this limitation, we plan to use a headless browser to render the SERPs and then parse them. A helpful library for this type of content extraction is Scriptor.%
\footnote{\url{https://github.com/webis-de/scriptor}}

The last and most pressing limitation, however, is that all URL captures and SERP contents must first be downloaded from the Internet Archive, which is restricted by both rate limits and network bandwidth. As estimated in Section~\ref{estimates}, currently~93\,\% of the SERPs still need to be fully downloaded from the Internet Archive and thus could not yet be parsed. Hence, more computational resources are required to use this extensive collection. In this regard, we will reach out to the Internet Archive whose privileged access to their infrastructure will allow for much faster compilation of the data.

\subsection{Contributing to the AQL}

There is an inherent boundary between search providers and researchers when using query logs. The AQL lowers this boundary by exploiting the web archival process of the Internet Archive. As we have described above, physical limitations, such as network speed, restrict the rate at which we can further grow the AQL. One way to overcome such limitations is to distribute computations across an open community, an approach that has been successfully employed in mathematics.%
\footnote{E.g., at the Great Internet Mersenne Prime Search: \url{https://mersenne.org/}}
We therefore open source our code to allow the community to contribute query and SERP parsers.

\section{Conclusion}
\label{conclusion}

The Archive Query Log provides an unparalleled academic resource. It consists of over 356~million queries, over 166~million SERPs, and over 1.7~billion results extracted from the SERPs, all coming from 550~search providers spanning 25~years. The AQL is the largest and most diverse query log ever publicly available. From an academic perspective, the AQL will enable researchers to tackle challenges in information retrieval that were not possible until now, ranging from the development of new models for retrieval, query suggestion or query prediction  to large-scale diachronic analyses of search engines; to name the most salient research avenues. Furthermore, our release plan for accessing the AQL minimizes the harm to society and will allow researchers to safely research the transparency and accountability of search engines while protecting user privacy.

In this paper, we have documented the initial version of the AQL (i.e., AQL-22). We have plans to release future versions of the AQL that will further expand the collection. First, we will continue to add to the long tail of search providers and continue our efforts to download and extract more data from the Internet Archive. Continuing to grow the types of data provided, the next version of the AQL will also include the content of web pages for each result in a SERP. Not least, we will investigate the training of large re-ranking models based on this data.

Altogether, the AQL is an exceedingly valuable resource for researchers and will enable advances in information retrieval research that were previously insurmountable due to the relatively low scale of query logs. Because of its scope, size, and diversity we consider the AQL a significant contribution to the community, and these dimensions will continue to grow as we build upon and expand future versions of the AQL.

\begin{acks}
This publication has received funding from the European Union's Horizon Europe research and innovation programme under grant agreement No 101070014 (OpenWebSearch.EU, \url{https://doi.org/10.3030/101070014})
\end{acks}

{
\bibliographystyle{ACM-Reference-Format}
\hbadness 10000
\bibliography{sigir23-archive-query-log-lit}
}

\end{document}